\documentclass{article}

\usepackage{microtype}
\usepackage{graphicx}
\usepackage{booktabs} 

\usepackage{bm, bbm}
\usepackage{subcaption, tikz}
\usepackage{multicol, multirow}
\usepackage[inline]{enumitem}

\usepackage[hyphens]{url}
\usepackage{hyperref}

\usepackage{float}

\usepackage[accepted]{icml2023}

\usepackage{amsmath}
\usepackage{amssymb}
\usepackage{mathtools}
\usepackage{amsthm}

\usepackage[capitalize,noabbrev]{cleveref}
\Crefname{ALC@unique}{Line}{Lines}

\theoremstyle{plain}
\newtheorem{theorem}{Theorem}[section]

\newtheorem{lemma}[theorem]{Lemma}

\newtheorem{example}[theorem]{Example}

\theoremstyle{definition}

\newtheorem{assumption}[theorem]{Assumption}
\theoremstyle{remark}

\usepackage[textsize=tiny]{todonotes}

\usepackage{xcolor}         
\definecolor{pku-red}{RGB}{139,0,18}

\newcommand{\calA}{\mathcal{A}}
\newcommand{\calB}{\mathcal{B}}

\newcommand{\calH}{\mathcal{H}}

\newcommand{\calV}{\mathcal{V}}

\newcommand{\bbE}{\mathbb{E}}

\newcommand{\bbR}{\mathbb{R}}

\newcommand{\vd}{\boldsymbol{d}}

\newcommand{\vv}{\boldsymbol{v}}

\icmltitlerunning{Learning to Bid in Repeated First-Price Auctions with Budgets}

\begin{document}

\twocolumn[
\icmltitle{Learning to Bid in Repeated First-Price Auctions with Budgets}



\icmlsetsymbol{equal}{*}

\begin{icmlauthorlist}
\icmlauthor{Qian Wang}{equal,cfcs}
\icmlauthor{Zongjun Yang}{equal,it}
\icmlauthor{Xiaotie Deng}{cfcs}
\icmlauthor{Yuqing Kong}{cfcs}
\end{icmlauthorlist}

\icmlaffiliation{cfcs}{Center on Frontiers of Computing Studies, Peking University, Beijing, China}
\icmlaffiliation{it}{School of Electronics Engineering and Computer Science, Peking University, Beijing, China}

\icmlcorrespondingauthor{Qian Wang}{charlie@pku.edu.cn}
\icmlcorrespondingauthor{Zongjun Yang}{allenyzj@stu.pku.edu.cn}

\icmlkeywords{Machine Learning, ICML}

\vskip 0.3in
]



\printAffiliationsAndNotice{\icmlEqualContribution} 

\begin{abstract}

Budget management strategies in repeated auctions have received growing attention in online advertising markets. However, previous work on budget management in online bidding mainly focused on second-price auctions. The rapid shift from second-price auctions to first-price auctions for online ads in recent years has motivated the challenging question of how to bid in repeated first-price auctions while controlling budgets.

In this work, we study the problem of learning in repeated first-price auctions with budgets. We design a dual-based algorithm that can achieve a near-optimal $\widetilde{O}(\sqrt{T})$ regret with full information feedback where the maximum competing bid is always revealed after each auction. We further consider the setting with one-sided information feedback where only the winning bid is revealed after each auction. We show that our modified algorithm can still achieve an $\widetilde{O}(\sqrt{T})$ regret with mild assumptions on the bidder's value distribution. Finally, we complement the theoretical results with numerical experiments to confirm the effectiveness of our budget management policy.

\end{abstract}

\section{Introduction}
Recent years have witnessed the explosive growth of the online advertising market. 
It is estimated that worldwide online advertising spending will reach 681 billion dollars in 2023, accounting for nearly 70\% of the entire advertising market spending \cite{adspending2022worldwide}. 
In practice, a huge amount of online ads are sold via real-time auctions implemented on advertising platforms and advertisers participate in such repeated online auctions to purchase advertising opportunities. 
An advertiser typically aims to maximize her cumulative payoffs during a specific time horizon (e.g., a day or a week) subject to a budget constraint, which reflects her monetary limit throughout this period. The presence of budgets constitutes an important operational challenge for online bidding algorithm design since there will be considerable financial losses whether depleting the budget too early or reaching the end of the period with unused funds. Therefore, budget management is a fundamental issue for designing practical online bidding algorithms. 


There has been a flourishing line of literature on budget management strategies in repeated second-price auctions~\cite{balseiro2019learning,balseiro2022best,chen2022dynamic}. 
However, a major industry-wide shift has occurred recently towards using first-price auctions as the preferred auction format of selling digital ads \cite{despotakis2021first}, as opposed to the earlier prevalent practice of using second-price auctions \cite{lucking2000vickrey,klemperer2004auctions,lucking2007pennies}. Google Ad Exchange, the largest online auction platform, announced its shift to the first-price auction
in September 2019~\cite{google2019fpa}. On the one hand, the difference in the nature of first-price auctions and second-price auctions implies that the algorithms proposed by the above work may not directly apply to the first-price setting. On the other hand, most previous work on repeated first-price auctions only considered bidding without budgets \cite{balseiro2019contextual,han2020optimal,han2020learning,badanidiyuru2021learning,zhang2022leveraging}. The shift thus leads to a pressing question of how should an advertiser bid in repeated first-price auctions to maximize the cumulative payoffs while controlling the expenditures. 

In this paper, we study the design of online bidding algorithms in repeated first-price auctions with budgets. 
We focus on a \textit{stochastic} setting where both the bidder's values and the maximum competing bids are \textit{i.i.d.} sampled over auctions. 
The goal is to optimize the bidder's expected cumulative rewards while keeping her budget constraint satisfied for any realization of values and competing bids. 
We provide online bidding algorithms for the bidder in two different feedback models: \begin{enumerate*}[label={(\arabic*)}]
    \item the \textit{full information feedback}, where the maximum competing bid is always revealed after each auction;
    \item the \textit{one-sided information feedback}, where only the winning bid is revealed after each auction.
\end{enumerate*}


\paragraph{Our Results.} Our main contribution is to propose two near-optimal bidding algorithms for each of the two feedback models. 
For the full information feedback, \Cref{algo: full} can achieve an $\widetilde{O}(\sqrt{T})$ regret where $T$ is the total number of auctions (\cref{thm: full}). 
For the the one-sided information feedback, \Cref{algo: one} can achieve an $\widetilde{O}(\sqrt{T})$ regret under mild assumptions on value distributions (\Cref{thm: one}).

Our algorithms follow a primal-dual framework and update a dual variable via online gradient descent to adjust the rate at which the bidder depletes her budget. The framework is similar to those used by previous work on second-price auctions \cite{balseiro2019learning,balseiro2022best,feng2022online,chen2022dynamic}. However, its application to repeated first-price auctions presents new challenges:
\begin{enumerate}
    \item First, due to the non-truthful nature of first-price auctions, we cannot compute the bid that maximizes the cost-adjusted reward without knowing the highest competing bid. One can only expect to maximize the objective in expectation by learning the hidden distribution of maximum competing bids. However, estimates using historical samples may not be sufficiently accurate. Even worse, with one-sided feedback, the bidder's observations are actually biased.
    \item Second, the dynamic update of the dual variable implies that the cost-adjusted reward function differs in different rounds. This causes failure in previous bandit algorithms like \citet{balseiro2019contextual} and \citet{han2020optimal}, as future rounds may suffer from exploitation when objectives are misaligned. 
\end{enumerate}

This work overcomes the two challenges and provides theoretical performance guarantees for our proposed algorithms. 
We start with the full information feedback and design \Cref{algo: full}, which sketches the high-level combination of online optimization methods and distribution estimation techniques. 
We then refine the algorithm for the one-sided information feedback, where we introduced value shading (i.e., scaling down the value by a factor) to align the objectives of different rounds so as to balance exploration and exploitation. 
We maintain a high-reward bid set for each shaded value and leverage the graph-feedback and partial-order properties in first-price auctions, which in essence follows the approach developed in \citet{han2020optimal}. 
As the bidder's observations are biased in this feedback model, the estimation errors are bounded via a martingale argument. The sum of estimation errors is shown to be upper bounded by $\widetilde{O}(\sqrt{T})$ with an assumption on the bidder's value distribution. In the experimental part, we demonstrate that our algorithms outperform those without budget management under various distributions in terms of the long-run average performance.


\subsection{Related Work} 

Learning in repeated auctions with constraints has been extensively studied in literature but most studies focused on only second-price auctions. 
\citet{balseiro2019learning} proposed an optimal online bidding algorithm known as adaptive pacing in repeated second-price auctions with budget constraints. 
\citet{balseiro2022best} extended the above work by relaxing some model assumptions and using a more general dual approximation scheme. 
\citet{feng2022online,golrezaei2021bidding} considered repeated second-price auctions with budget and return-on-spend (RoS) constraints.
\citet{chen2022dynamic} studied another important class of budget management strategies, called throttling,
and proposed a near-optimal throttling algorithm for repeated second-price auctions with budgets. 
Our work considers budget management in repeated first-price auctions instead but we use some similar techniques to those in the above papers.

For repeated first-price auctions, most previous work only considered bidding without constraints. \citet{balseiro2019contextual} first considered learning in repeated first-price auctions with binary feedback where the bidder only knows whether she wins or not.
They adopted a cross-learning approach to achieve an $\widetilde{O}(T^{2/3})$ regret and
showed that the lower bound on regret
is $\Omega(T^{2/3})$. 
\citet{han2020optimal} considered repeated first-price auctions with one-sided feedback, where the winning bid is revealed after each auction. 
They leveraged the graph-feedback and partial-order properties in first-price auctions to achieve an $\widetilde{O}(\sqrt{T})$ regret
and proved that the lower bound on regret
is $\Omega(\sqrt{T})$ even under full information feedback where the maximum competing bid is always revealed after each auction. \citet{han2020learning} considered the setting
without the \textit{i.i.d.} assumption of the maximum competing bids
and achieved an $\widetilde{O}(\sqrt{T})$ regret when competing with the set of all Lipschitz bidding strategies. \citet{badanidiyuru2021learning} studied the contextual first-price auctions where the values and the maximum competing bids depend on a public context. 
\citet{zhang2022leveraging} studied the setting where the bidder has access to some hint relevant to the maximum competing bid.
The main difference between the above papers and ours is that the bidder has a budget constraint in our model.

\citet{ai2022no} also studied no-regret learning in repeated first-prices auctions with budgets. However, their model additionally involves a discount factor $\gamma < 1$ in the objective function and the ``optimal'' strategy is defined with respect to this variant problem. 
Their algorithm will abort if $\gamma = 1$ so our results are not directly comparable with theirs.


\citet{balseiro2022contextual} studied the equilibrium biding strategies for first-price auctions with budgets. Their value-pacing-based strategies are similar to our second algorithm, but we are investigating a dynamic setting from the view of a single budget-constrained bidder. The equilibrium characterization of the first-price market cannot provide regret guarantees for dynamic bidding, especially considering the learning process with respect to different feedback models.


\section{Model and Benchmark}

In this work, we consider the problem of online learning in the first-price auction market. We focus on a single bidder in a large population of bidders during a time horizon $T$. 

In each round $t = 1, \ldots, T$, there is an available ad slot auctioned by the seller (e.g. an advertising platform). 
The bidder receives a private value $v_t\in [0, \bar{v}]$, and then submits a bid $b_t \in \bbR_+$ based on $v_t$ and all historical observations available to her. 
We denote the maximum bid of all other bidders by $d_t\in \bbR_+$. 
The auction outcome depends on the comparison between $b_t$ and $d_t$. 
Let $x_{t} \coloneqq \bm{1}\left\{b_{t} \geq d_{t}\right\}$ be the binary variable indicating whether the bidder wins the ad slot at round $t$. 
Here we assume that ties are broken in favor of the bidder we concern to simplify exposition. We note that this choice is arbitrary and by no means a limitation of our approach.
Let $r_{t} \coloneqq x_{t}(v_{t} - b_{t})$ be her reward and let $c_{t} \coloneqq x_{t}b_{t}$ be the corresponding cost for a first-price auction.
As usual, we use the bold symbol $\vv$ without subscript $t$ to denote the vector $(v_{1}, \ldots, v_{T})$; the same goes for other variables in the present paper.

The bidder has a budget $B$ that limits the payments she can make over $T$ rounds of auctions, and her maximum expenditure rate is denoted by $\rho \coloneqq B/T$. 
We assume that $\rho \in (0, \bar{v}]$; otherwise, it becomes a problem without constraints as the bidder would never deplete her budget.

We consider a \textit{stochastic} setting where $v_t$ is \textit{i.i.d.} sampled from a distribution $F$ and $d_t$ is \textit{i.i.d.} sampled from a distribution $G$. 
The latter assumption follows from the standard mean-field approximation \cite{iyer2014mean,balseiro2015repeated} and is a common practice in literature. 
The main rationale behind this assumption is that when the number of other bidders is large, on average their valuations and bidding strategies are static over time. 
Note that both $F$ and $G$ are unknown to the bidder.

\paragraph{Information structure.} 
In repeated first-price auctions, the bidder can receive different feedback after each round depending on the information released by the seller. 
In particular, the ability of this bidder to observe the maximum competing bid $d_t$ varies to the information structure. In this paper, we investigate two different information structures:
\begin{enumerate}
    \item \textit{Full information feedback.} 
    The bidder can observe the maximum competing bid $d_t$ at the end of each round $t$. 
    This information structure makes sense in many current online auction platforms. 
    For example, in the Google Ad Exchange, at the end of an auction, bidders will receive back the minimum value they would have had to bid to win the auction, whether they lose or win \cite{google2022adshelp}.
    \item \textit{One-sided information feedback.} The bidder can observe the maximum competing bid $d_t$ only if she loses the auction. 
    Thus, the feedback available to her includes $\bm{1}\{b_t\geq d_t\}$ and $d_t\bm{1}\{b_t < d_t\}$.
    This can be viewed as an informational version of the \textit{winner’s curse} \cite{capen1971competitive} where the winner learns less information. And this is a common feedback model in previous studies on repeated first-price auctions \cite{esponda2008information,han2020optimal,ai2022no}. Compared to the full information feedback, the one-sided information feedback is more complicated to deal with as the bidder can observe less information.
\end{enumerate}

We denote the historical observations available to the bidder before submitting a bid in round $t$ by $\calH_t$. For the full information feedback, we define
\begin{align*}
    \calH^F_t \coloneqq \left(v_{s}, x_{s}, d_{s}\right)_{s=1}^{t-1}.
\end{align*}
At the end of each round $s$, the bidder can append a tuple $(v_{s}, x_{s}, d_{s})$ to the available history. For the one-sided information feedback, we define
\begin{align*}
    \calH^O_t \coloneqq \left(v_{s}, x_{s}, (1-x_s)d_{s}\right)_{s=1}^{t-1}.
\end{align*}
At the end of each round $s$, the bidder knows her value and whether she wins, but the winner can only observe $(1-x_s)d_s = 0$.

\paragraph{Bidding strategy and regret.} A bidding strategy maps $(\calH_t, v_t)$ to a (possibly random) bid $b_t$ for each $t$. We say $\pi$ is \textit{budget feasible} if it generates expenditures that are constrained by the budget for any realizations of values and maximum competing bids, i.e. $\forall \vv, \vd$, 
\begin{align*}
    \sum^T_{t = 1} c^{\pi}_t = \sum^T_{t = 1} \bm{1}\left\{b^{\pi}_t \geq d_t\right\} b^{\pi}_t \leq B = \rho T.
\end{align*}
We denote by $\Pi_0$ the set of all budget feasible strategies. For a strategy $\pi \in \Pi_0$, we denote by $R(\pi)$ the performance of $\pi$, defined as follows:
\begin{align*}
    R(\pi) = & \bbE^{\pi}_{\vv, \vd} \left[\sum_{t=1}^T r^{\pi}_t\right] \\
    = & \bbE^{\pi}_{\vv, \vd} \left[\sum_{t=1}^T \bm{1}\left\{b^{\pi}_{t} \geq d_{t}\right\}\left(v_t - b_t^{\pi}\right)\right],
\end{align*}
where the expectation is taken with respect to the values, the maximum competing bids and any possible randomness embedded in the strategy. The bidder's optimization problem can be written as
\begin{equation}
\label{eqn: problem}
    \begin{aligned}
    \max_{\pi} & \ \bbE^{\pi}_{\vv, \vd} \left[\sum_{t=1}^T \bm{1}\left\{b^{\pi}_{t} \geq d_{t}\right\}\left(v_t - b_t^{\pi}\right)\right] \\
    \text{s.t. } & \ \sum_{t=1}^T \bm{1}\left\{b^{\pi}_{t} \geq d_{t}\right\}b^{\pi}_t \leq \rho T, \ \forall \vv, \vd.
    \end{aligned}
\end{equation}

The regret of the bidder is defined to be the difference in the expected cumulative rewards of the bidder's strategy and the optimal budget feasible bidding strategy, which has the perfect knowledge of $F$ and $G$:
\begin{align*}
    Reg(\pi) = \max_{\pi'\in \Pi_0} R(\pi') - R(\pi).
\end{align*}





\section{Bidding Algorithms and Analysis}

In this section, we first design and analyze an algorithm for Problem~\eqref{eqn: problem} with full information feedback, which reveals our high-level idea on budget management in repeated first-price auctions. We prove that the algorithm can achieve an $\widetilde{O}(\sqrt{T})$ regret. Then we modify our algorithm to accommodate the setting with only one-sided information feedback by using value shading and leveraging a special partial order property possessed by first-price auctions. The modified algorithm can still achieve an $\widetilde{O}(\sqrt{T})$ regret under mild assumptions. All omitted proofs in this section can be found in the appendix.

\subsection{Full Information Feedback}
\label{sec: full}

Our bidding algorithm for full information feedback is depicted in \Cref{algo: full}. The bidder first conducts a one-round exploration to make an appropriate initialization (\Cref{algo: full initialization}). After observing the value $v_t$ in each round $t=2, \ldots, T$, the bidder constructs estimates of empirical rewards and costs based on all historical observations and submits a bid that maximizes a cost-adjusted reward (\Cref{algo: full value,algo: full estimate,algo: full bid}). Then the bidder updates $\lambda_t$ using the empirical cost of the submitted bid $\widetilde{c}_t(b)$ and her average budget $\rho$ (\Cref{algo: full update}). The variable $\lambda_t$ plays a key role in adjusting the pace at which the bidder depletes her budget. By the choice of $b_t$, we must have $b_t \leq v_t/(1+\lambda_t)$. If the bidder bids too high in past rounds, $\lambda_t$ tends to be larger, thereby controlling the bids in future rounds. 

\begin{algorithm}[t]
    \caption{Bidding Algorithm for First-Price Auctions with Budgets under Full Information Feedback}
    \label{algo: full}
    \begin{algorithmic}[1]
    \setcounter{ALC@unique}{0}
        \STATE {\bfseries Input:} Time horizon $T$; budget $B=\rho T$; update step $\epsilon > 0$; failure probability $\delta > 0$.
        \STATE {\bfseries Initialization:} The bidder bids $b_1 = 0$ and set $B_2 = B, \lambda_2 = 0$. \label{algo: full initialization}
        \FOR{$t\in\{2,\cdots,T\}$}
            \STATE The bidder receives the value $v_t \in [0,\bar{v}]$. \label{algo: full value}
            \STATE The bidder estimates the rewards and costs: \label{algo: full estimate}
            \begin{align}
                \widetilde{r}_t(v_t, b) = & \frac{1}{t-1}\sum_{s = 1}^{t-1}\bm{1}\left\{b \geq d_s\right\}\left(v_t - b\right), \label{eqn: estimate r full}\\
                \widetilde{c}_t(b) = & \frac{1}{t-1}\sum_{s = 1}^{t-1}\bm{1}\left\{b \geq d_s\right\}b. \label{eqn: estimate c full}
            \end{align}
            \STATE The bidder submits a bid: \label{algo: full bid}
            \begin{align}
                b_t \in \arg\max_b \left(\widetilde{r}_t(v_t, b) - \lambda_t \widetilde{c}_t(b)\right). \label{eqn: best b full}
            \end{align}
            (Taking the smallest if there are ties.)
            \STATE The bidder updates the parameter \label{algo: full update}
            \begin{align}
                \lambda_{t+1} = {\rm Proj}_{\lambda > 0} \left(\lambda_t - \epsilon\left(\rho - \widetilde{c}_t(b_t)\right) \right).
            \end{align}
            \STATE The bidder observes the maximum competing bid $d_t$.
            \STATE The bidder update the remaining budget 
            \begin{align}
                B_{t+1} = B_t - c_t.
            \end{align}
            \IF{$B_{t+1} < \bar{v}$}
                \STATE \textbf{break}
            \ENDIF
        \ENDFOR
    \end{algorithmic}
\end{algorithm}

To provide more intuition on the choice of $b_t$ and the update procedure of $\lambda_t$, we consider an alternative optimization problem with a soft budget constraint:
\begin{equation}
\label{eqn: soft problem}
    \begin{aligned}
    \max_{\pi} & \ \bbE^{\pi}_{\vv, \vd} \left[\sum_{t=1}^T \bm{1}\left\{b^{\pi}_{t} \geq d_{t}\right\}\left(v_t - b_t^{\pi}\right)\right] \\
    \text{s.t. } & \ \bbE^{\pi}_{\vv, \vd} \left[\sum_{t=1}^T \bm{1}\left\{b^{\pi}_{t} \geq d_{t}\right\}b^{\pi}_t\right] \leq \rho T.
    \end{aligned}
\end{equation}
The Lagrangian dual objective of Problem~\eqref{eqn: soft problem} is
\begin{align*}
    & \bbE^{\pi}_{\vv, \vd} \left[\sum_{t=1}^T \left(\bm{1}\left\{b^{\pi}_{t} \geq d_{t}\right\}\left(v_t - (1+\lambda)b_t^{\pi}\right) + \lambda\rho\right)\right] \\
    = & \sum_{t=1}^T \left(\bbE^{\pi}_{\calH_t, v_t} \Big[\left(v_t - (1+\lambda)b_t^{\pi}\right)G\left(b^{\pi}_{t}\right)\Big] + \lambda\rho \right),
\end{align*}
where the equality holds since $b_t^{\pi}$ is independent of $d_t$ as well as other future values and maximum competing bids. For a fixed $\lambda$, the dual objective is maximized by bidding $b_t \in \arg\max_b (v_t - (1+\lambda)b)G(b)$, which is irrelevant to the historical observations $\calH_t$. 

We denote by $\Pi_1$ the set of all strategies that satisfy the soft budget constraint in Problem~\eqref{eqn: soft problem}. By weak duality, we have
\begin{align}
    & \max_{\pi \in \Pi_1} R(\pi) \nonumber \\
    \leq & \min_{\lambda\geq 0} T \left(\bbE_{v} \left[\max_b \left(v - (1+\lambda)b\right)G\left(b\right)\right] + \lambda\rho \right). \label{eqn: soft problem upper bound}
\end{align}
Our algorithm adopts an online gradient descent scheme to approximate the right hand side of \eqref{eqn: soft problem upper bound}, which is also an upper bound on the optimal value of Problem~\eqref{eqn: problem} since $\Pi_0 \subseteq \Pi_1$. A crucial challenge here is that the bidder does not know the prior distribution $G$ so she cannot calculate the exact maximum point of $(v - (1+\lambda)b)G(b)$. To deal with this issue, we adopt the distribution estimation method and use $\widetilde{r}_t(v_t, b), \widetilde{c}_t(b)$ in place of $(v-b)G(b), bG(b)$. As $t$ grows, the estimates become more accurate. The following lemma shows that $\forall v_t, b$, $\widetilde{r}_t(v_t, b)$ and $\widetilde{c}_t(b)$ are good estimates of $r(v_t, b) \coloneqq (v_t - b)G(b)$ and $c(b) \coloneqq bG(b)$ respectively. 
\begin{lemma}
\label{lem: full dkw}
Under \Cref{algo: full}, with probability at least $1 - \delta$, we have for all $t \geq 2$ and $b \leq \bar{v}$,
\begin{align}
    |\widetilde{r}_t(v_t, b) - r(v_t, b)| \leq & \bar{v}\cdot \sqrt{\frac{\ln \left(2T/\delta\right)}{2(t-1)}}, \label{eqn: full dkw r} \\
    |\widetilde{c}_t(b) - c(b)| \leq & \bar{v}\cdot \sqrt{\frac{\ln \left(2T/\delta\right)}{2(t-1)}}. \label{eqn: full dkw c}
\end{align}
\end{lemma}

\begin{theorem}
\label{thm: full}
For repeated first-price auctions with budget constraints and full information feedback, \Cref{algo: full} can achieve
\begin{align*}
    Reg(\pi) = O\left(\sqrt{T\ln T}\right).
\end{align*}
\end{theorem}

In the proof of \Cref{thm: full}, we first we perform a standard analysis of the online gradient descent method to show the sequence of $\lambda_t$ is not much worse than a hindsight $\lambda$ with respect to gain function $h_t(\lambda_t) = \lambda_t\left(\widetilde{c}_t(b) - \rho\right)$. This, together with \Cref{lem: full dkw} implies that with high probability, the bid $b_t$ chosen in each round is close to the bid chosen by the best strategy for Problem~\eqref{eqn: soft problem}, i.e., $\arg\max_b (v_t - (1+\lambda^*)b)G(b)$ where $\lambda^*$ is the optimal dual variable. Finally, the proof is concluded by showing that the time at which the budget is depleted under \Cref{algo: full} is close to $T$.

\paragraph{Lower bound.} As previous work has proved, the lower bound on regret for this problem is $\Omega(\sqrt{T})$ even without constraints (i.e., $\rho = \bar{v}$).

\begin{lemma}[\citet{han2020optimal}]
For repeated first-price auctions and full information feedback, there exists a positive constant $C > 0$ independent of $T$ such that
\begin{align}
    \inf_{\pi} \sup_{F, G} Reg(\pi) \geq C\sqrt{T}.
\end{align}
\end{lemma}

This previous result implies that our algorithm can achieve a near-optimal learning performance in first-price auctions with budget constraints.

\paragraph{Discretization.} In \Cref{algo: full}, \Cref{algo: full estimate} requires estimating rewards and costs for all possible bids but the bid space might be continuous. In practice, we can resolve this issue by a simple discretization, which will cause little performance degradation. Let $\calB = \{b^1, \cdots, b^K\}$ with $b^k = (k-1)/K\cdot \bar{v}$. We then change \Cref{algo: full bid} to that the bidder submits a bid
\begin{align}
\label{eqn: best b full quantize}
    b_t \in \arg\max_{b\in \calB} \left(\widetilde{r}_t(v_t, b) - \lambda_t \widetilde{c}_t(b)\right). 
\end{align}
When $K = \Omega(\sqrt{T})$, the discretization error is of order $O(\sqrt{T})$. In the next subsection, we will discretize both values and bids in \Cref{algo: one} and more formally analyze the additional regret caused by the discretization.
\subsection{One-sided Information Feedback}
\label{sec: one}

This subsection provides a modified algorithm for the scenario with one-sided information feedback. We show an $\widetilde{O}(\sqrt{T})$ regret can still be achieved with an assumption on the bidder's value distribution.

As the bidder can only observe the highest competing bid $d_s$ after losing at round $s$ under one-sided information feedback, she can no longer estimate the expected rewards and costs in each round using all past rounds as in \Cref{algo: full}. Specifically, given value $v_t$ and bid $b$, if $b < b_s$ and $b_s \geq d_s$, she cannot determine $\bm{1}\{b \geq d_s\}$, so that she cannot calculate $\widetilde{r}_t(v_t, b), \widetilde{c}_t(b)$ as in \eqref{eqn: estimate r full}, \eqref{eqn: estimate c full}. Therefore, we need new estimators for the expected rewards and costs. 

For this purpose, we first discretize the value space into a set of size $M$, $\calV = [v^1, \cdots, v^M]$ with $v^m = (m-1)/M \cdot \bar{v}$, and the bid space into a set of size $K$, $\calB = \{b^1, \cdots, b^K\}$ with $b^k = (k-1)/K \cdot \bar{v}$. Then, we denote by $n_t^k$ the number of observed bids lower than $b^k$ before round $t$:
\begin{align}
    n_t^k \coloneqq \sum_{s=1}^{t-1} \bm{1}\left\{b_s \leq b^k\right\}. \label{eqn: count one}
\end{align}
Given $v^m$ and $b^k$, we define two estimators under one-sided information feedback as
\begin{align}
    \widetilde{r}_t(v^m, b^k) = &\frac{1}{n_t^k}\sum_{s = 1}^{t-1}\bm{1}\left\{b_s \leq b^k\right\}\bm{1}\left\{b^k \geq d_s\right\}\left(v^m - b^k\right), \label{eqn: estimate r one} \\
    \widetilde{c}_t(b^k) = & \frac{1}{n_t^k}\sum_{s = 1}^{t-1}\bm{1}\left\{b_s \leq b^k\right\} \bm{1}\left\{b^k \geq d_s\right\}b^k. \label{eqn: estimate c one}
\end{align}
Note that the equations \eqref{eqn: estimate r one} and \eqref{eqn: estimate c one} are measurable with respect to the available history $\calH_t^O$. If $b_s \geq d_s$, we have $\bm{1}\left\{b_s \leq b^k\right\}\bm{1}\left\{b^k \geq d_s\right\} = \bm{1}\left\{b_s \leq b^k\right\}$; if $b_s \leq d_s$, the bidder can observe the exact $d_s$ to determine $\bm{1}\left\{b^k \geq d_s\right\}$.

Since $(d_s)_{s=1}^{t-1}$ are not mutually independent conditioned on $(b_s)_{s=1}^{t-1}$, $\widetilde{r}_t(v^m, b^k)$ and $\widetilde{c}_t(b^k)$ are actually not unbiased estimators. However, we can still prove via a martingale argument that they approximate well to the expected reward $r(v^m, b^k) = (v^m-b^k)G(b^k)$ and the expected cost $c(b^k) = b^kG(b^k)$ with high probability, which is formalized as the following lemma.

\begin{lemma}
\label{lem: one azuma}
Under \Cref{algo: one}, with probability at least $1-\delta$, we have $\forall t\geq 2, m\in [M], k\in [K]$,
\begin{align}
    |\widetilde{r}_t(v^m, b^k) - r(v^m, b^k)| \leq & \bar{v}\cdot \sqrt{\frac{4 \ln{T}\ln{\left(KT/\delta\right)}}{n_t^k}}, \\
    |\widetilde{c}_t(b^k) - c(b^k)| \leq & \bar{v}\cdot\sqrt{\frac{4 \ln{T}\ln{\left(KT/\delta\right)}}{n_t^k}}.
\end{align}

\end{lemma}

By \Cref{lem: one azuma}, we know that the more bids that are lower than $b^k$, the more accurate the estimation of $r(v^m, b^k)$ and $c(b^k)$. However, bidding low in order to benefit future estimates may cause great loss in the current round. The existence of the budget constraint further increases the difficulty of balancing present and future rewards.

\begin{algorithm}[p]
\caption{Bidding Algorithm for First-Price Auctions with Budgets under One-Sided Information Feedback}\label{algo: one}
\begin{algorithmic}[1]
\setcounter{ALC@unique}{0}
        \STATE {\bfseries Input:} Time horizon $T$; budget $B=\rho T$; 
        value set $\calV = [v^1, \cdots, v^M]$ with $v^m = (m-1)/M \cdot \bar{v}$; 
        bid set $\calB = \{b^1, \cdots, b^K\}$ with $b^k = (k-1)/K \cdot \bar{v}$; 
        update step $\epsilon > 0$; 
        failure probability $\delta\in (0,1)$.
        \STATE {\bfseries Initialization:} The bidder bids $b_1 = 0$ and set $B_2 = B, \lambda_2 = 0$. Set $\calB_0^m \gets \calB$ for each $v_m \in \calV$. \label{algo: one initialization}
	\FOR{$t\in\{2,\cdots,T\}$}
        \STATE The bidder receives the value $v_t \in [0,1]$;
        \STATE The bidder counts the observations by \eqref{eqn: count one}, and estimates the rewards and costs  by \eqref{eqn: estimate r one} and \eqref{eqn: estimate c one}. \label{algo: one estimate}
        \FOR{$m\in\{1,2,\cdots,M\}$}
        \STATE The bidder eliminates bids by: \label{algo: one partial order}
        \begin{align}
            \calB_{t-1}^m = \left\{b^k\in \calB_{t-1}^m: b^k \geq \max_{s < m} \inf \calB_t^{s}\right\}.
        \end{align}
        \STATE The bidder computes the confidence bound:
        \begin{align}
            w_t^m = \bar{v}\cdot \sqrt{\frac{4\ln{T}\log(KT/\delta)}{N_t^m}},
        \end{align}
        where $N_t^m = \min_{b^k\in \calB_{t-1}^m} n_t^k$.
        \STATE The bidder eliminates bids by: \label{algo: one high reward bid}
            \begin{align}
                \calB_t^m \gets \Big\{& b^k\in  \calB_{t-1}^m: \widetilde{r}_t(v^m, b^k) \nonumber \\
                \ge & \max_{b^{k'} \in \calB_{t-1}^m}\widetilde{r}_t(v^m, b^{k'}) - 2 w_t^m \Big\}. 
            \end{align}
        \ENDFOR
        \STATE The bidder chooses \label{algo: one value shading}
        \begin{align}
            v^{m(t)} = \max\{u\in \calV: u \leq v_t/(1+\lambda_t)\}.
        \end{align}
        \STATE The bidder submits a bid $b_t = \inf \calB_{t}^{m(t)}$; \label{algo: one bid}
            \STATE The bidder updates the parameter \label{algo: one update}
            \begin{align}
                \lambda_{t+1} = {\rm Proj}_{\lambda > 0} \left(\lambda_t - \epsilon\left(\rho - \widetilde{c}_t(b_t)\right) \right).
            \end{align}
        \STATE The bidder observes $x_t$ and $(1-x_t)d_t$;
        \STATE The bidder update the remaining budget 
        \begin{align}
            B_{t+1} = B_t - c_t.
        \end{align}
        \IF{$B_{t+1} < \bar{v}$}
            \STATE \textbf{break}
        \ENDIF
	\ENDFOR
\end{algorithmic}
\end{algorithm}

We depict our modified algorithm for one-sided information feedback in \cref{algo: one}. The main difference with \Cref{algo: full} is that the bidder maintains an active set of high reward bids for each $m\in [M]$. After observing the value $v_t$ in round $t$, the bidder shades it by $(1+\lambda_t)$ (\Cref{algo: one value shading}) and rounds it to $v^{m(t)}\in \calV$. Then the bidder submits a bid among the active bid set $B^{m(t)}_t$ (\Cref{algo: one bid}).
The value-shading step essentially aligns the objectives of different rounds. 
Observe that
\begin{align*}
    r(v_t, b^k) - \lambda_t c(b^k) = & \left(v_t - \left(1+\lambda_t\right)b^k\right)G(b^k) \\
    = & (1+\lambda_t) \left(\frac{v_t}{1+\lambda_t} - b^k\right)G(b^k) \\
    = & (1+\lambda_t) \cdot r(v_t/(1+\lambda_t), b^k).
\end{align*}
Thus, with $v^m\approx v_t/(1+\lambda_t)$, maximizing $r(v_t, b) - \lambda_t c(b)$ is approximately equivalent to maximizing the reward $r(v^m, b)$ as if the bidder is participating in a first-price auction without any budget constraint and the value is $v^m$. 

Instead of submitting the bid with the highest estimated reward $\widetilde{r}_t({v^m, b^k})$, the algorithm chooses the smallest bid in $\calB_t^m$ in order to make future estimates as accurate as possible. In addition to filtering bids according to the expected rewards and confidence bounds (\Cref{algo: one high reward bid}), the algorithm also eliminates some small bids from the active sets in \Cref{algo: one partial order}. This elimination increases $N_t^m$ so that the bidder can use smaller confidence bounds to prune the active sets. As long as the best bid for a value $v^m$ remains in the active set $\calB^m_t$, the elimination can further control the regret.

To prove that the best bids are not eliminated, we leverage a special partial order property of first price auctions
, i.e., 
$b^*(v) = \arg\max_b (v-b)G(b)$ is non-decreasing in $v$. In particular, for our scenario with discretization, $\widetilde{b}(v) = \arg\max_{b\in \calB} (v-b)G(b)$ is non-decreasing in $v$. The property guarantees that with high probability, $\widetilde{b}(v^m)$, which approximately maximizes $r(v^m, b)$, is not eliminated from the active set $B^m_{t-1}$ by \Cref{algo: one partial order}. (See \Cref{lem: one elimination}.)



\begin{assumption}
\label{asm: main}
The cumulative probability distribution $F$ is continuous with bounded density function $f$ satisfying $0 < \underline{f} < f(v) < \overline{f} < \infty$ for $v\in [0, \bar{v}]$.
\end{assumption}

\Cref{asm: main} is a technical assumption required by our analysis. We note that the existence of positive bounds on the density function is a common assumption in various learning problems. For example, \citet{balseiro2019learning} took it as one of the sufficient conditions for the strong convexity of the dual objective function.

\begin{lemma}
\label{lem: main}
Suppose that \Cref{asm: main} holds. We have
\begin{align*}
    \bbE_{\vv, \vd} \left[\sum_{t=2}^T \sqrt{1\big/N_t^{m(t)}}\right] \leq \widetilde{O}\left(\sqrt{T}\right).
\end{align*}
\end{lemma}

For full information feedback, the estimation error in round $t$ is $O(\sqrt{\ln T/(t-1)})$ by \Cref{lem: full dkw} so that we are able to control the sum of errors within $\widetilde{O}(\sqrt{T})$. \Cref{lem: main} establishes that for one-sided information feedback, we can get a similar result given \Cref{asm: main} holds, which constitutes a key part of the proof of the following \Cref{thm: one}.

\begin{theorem}
\label{thm: one}
Suppose that \Cref{asm: main} holds. For repeated first-price auctions with budget constraints and one-sided information feedback, there exists constants $C_1, C_2, C_3$, such that \Cref{algo: one} can achieve
\begin{align*}
    Reg(\pi) \leq C_1\sqrt{T\ln{(KT^2)}}\ln{T} + C_2\frac{T}{M} + C_3\frac{T}{K}.
\end{align*}
\end{theorem}

Particularly, when choosing $K = M = O(\sqrt{T})$, we obtain that $Reg(\pi) \leq \widetilde{O}(\sqrt{T})$ when $T$ is sufficiently large. We note that to prove \Cref{thm: one}, \Cref{asm: main} is sufficient but may not be necessary. In \Cref{sec: exp}, we run numerical experiments with further discussion.
\section{Experiments}
\label{sec: exp}

\begin{figure*}[h] 
    \centering
    \begin{subfigure}{0.33\textwidth}
	\centering
	\includegraphics[width=\linewidth]{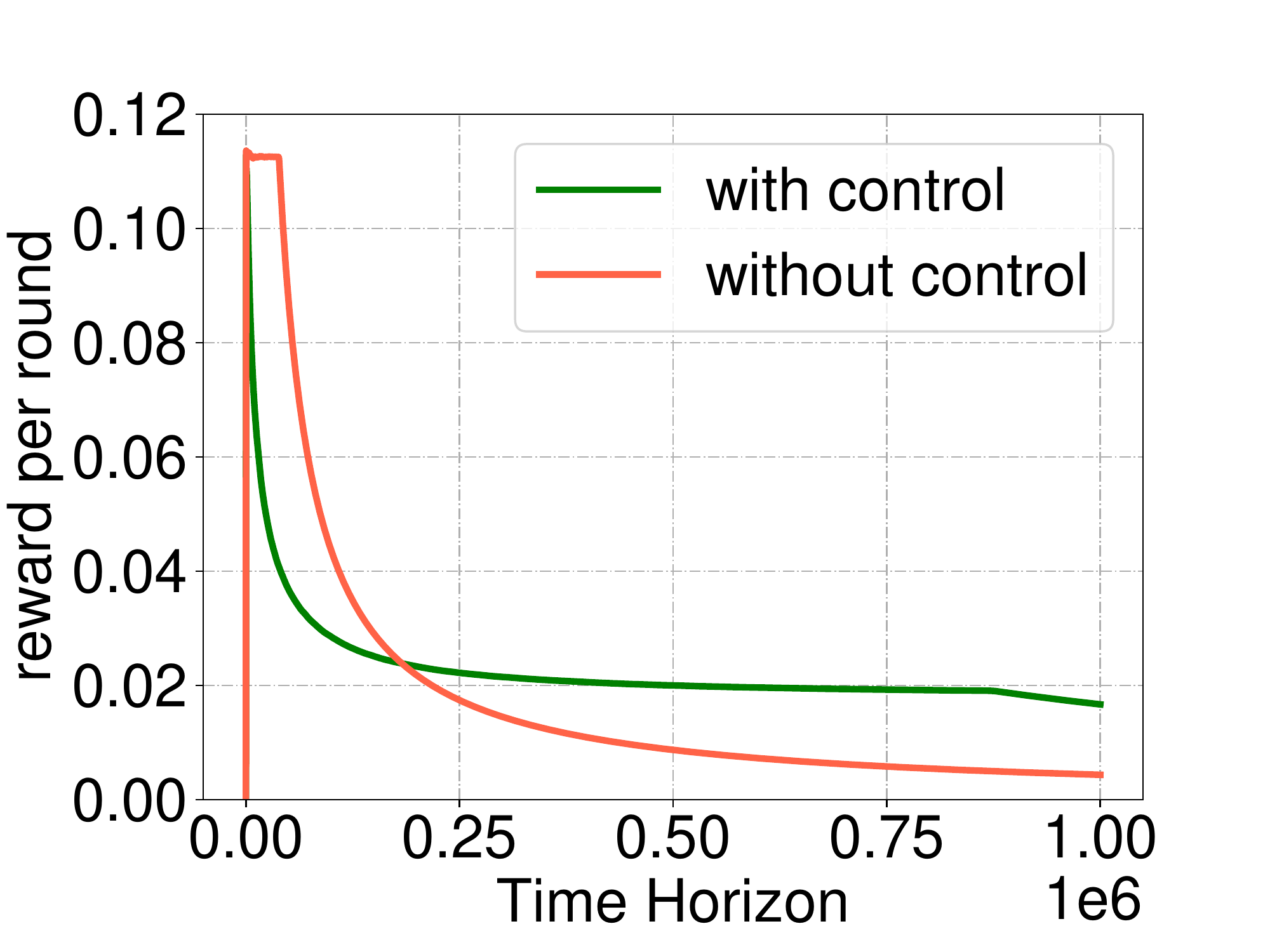}
	\caption{Normal $v_t$ , full feedback.}
	\end{subfigure}
    \begin{subfigure}{0.33\textwidth}
	\centering
	\includegraphics[width=\linewidth]{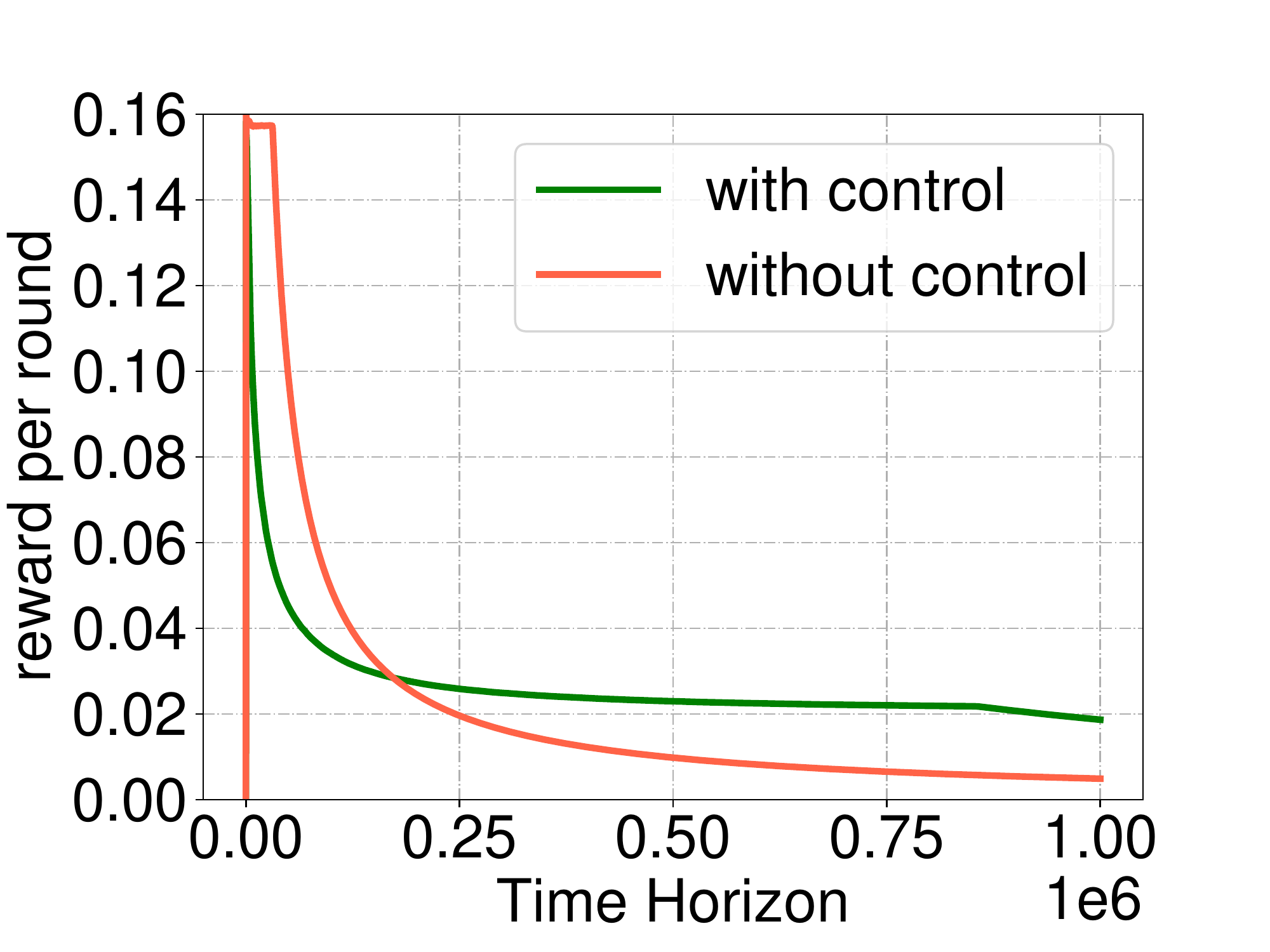}
	\caption{Log-normal $v_t$, full feedback.}
	\end{subfigure}
 \begin{subfigure}{0.33\textwidth}
	\centering
	\includegraphics[width=\linewidth]{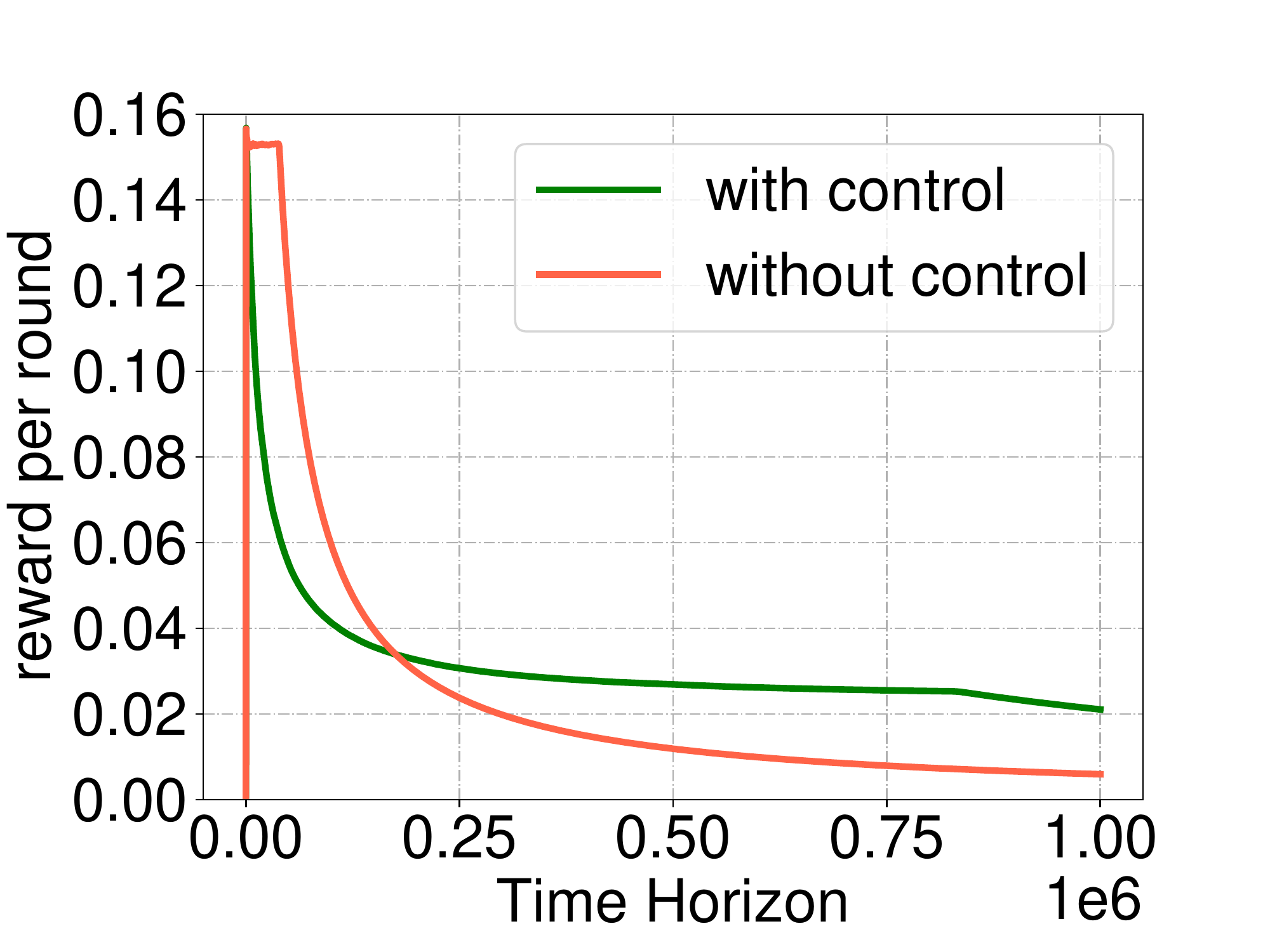}
	\caption{Uniform $v_t$, full feedback.}
	\end{subfigure}
 \begin{subfigure}{0.33\textwidth}
	\centering
	\includegraphics[width=\linewidth]{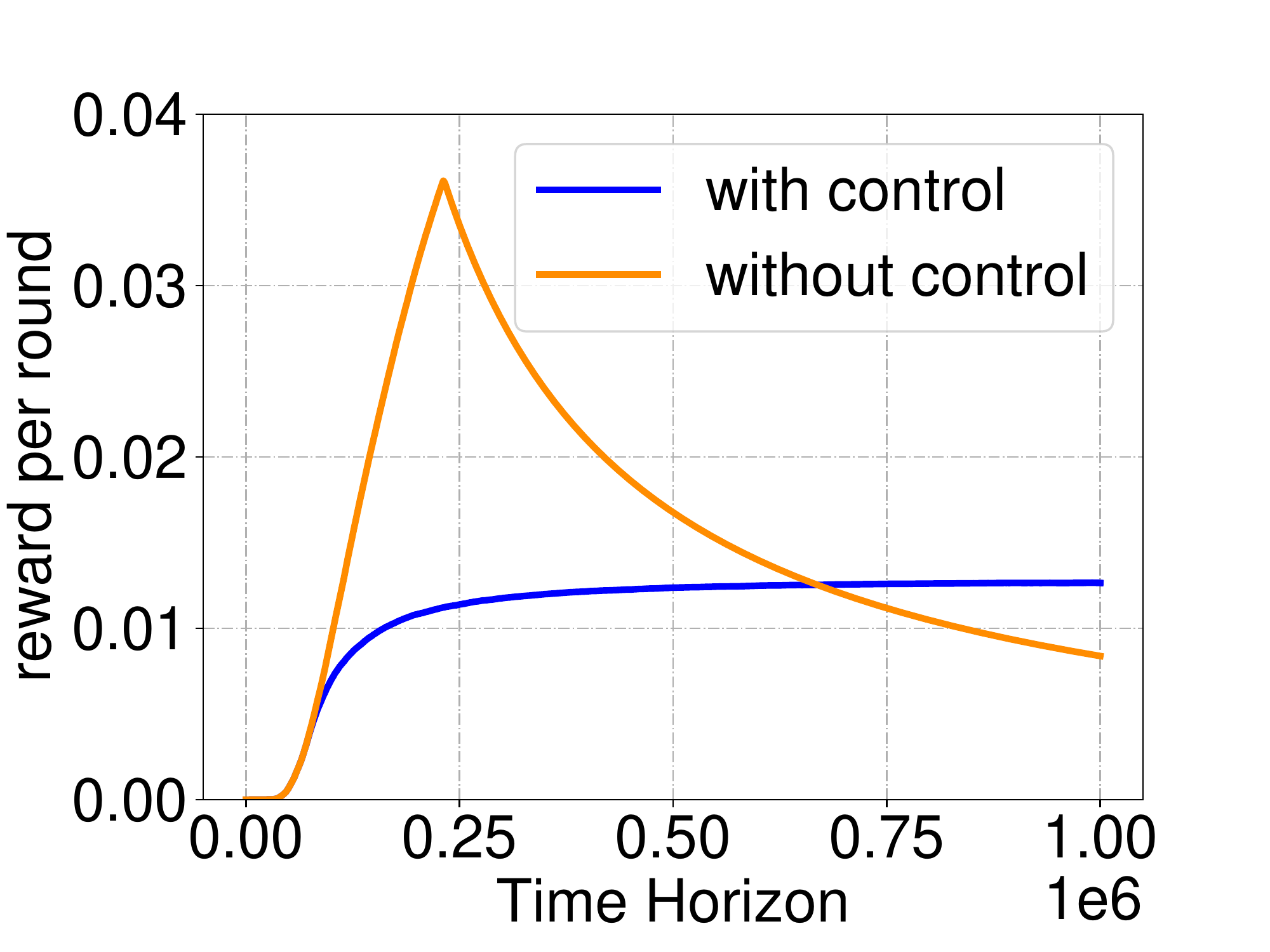}
	\caption{Normal $v_t$, one-sided feedback.}
	\end{subfigure}
 \begin{subfigure}{0.33\textwidth}
	\centering
	\includegraphics[width=\linewidth]{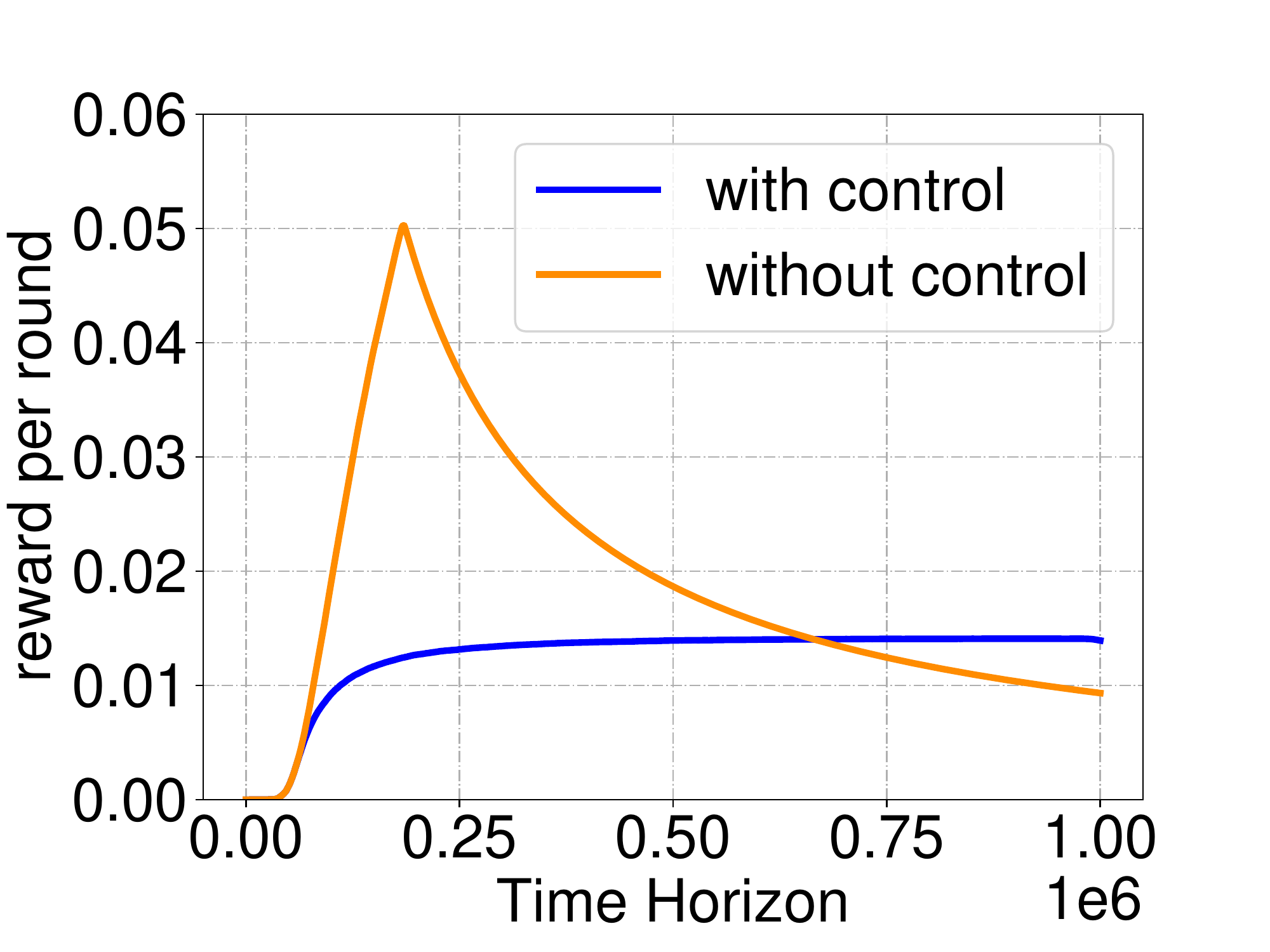}
	\caption{Log-normal $v_t$, one-sided feedback.}
	\end{subfigure}
 \begin{subfigure}{0.33\textwidth}
	\centering
	\includegraphics[width=\linewidth]{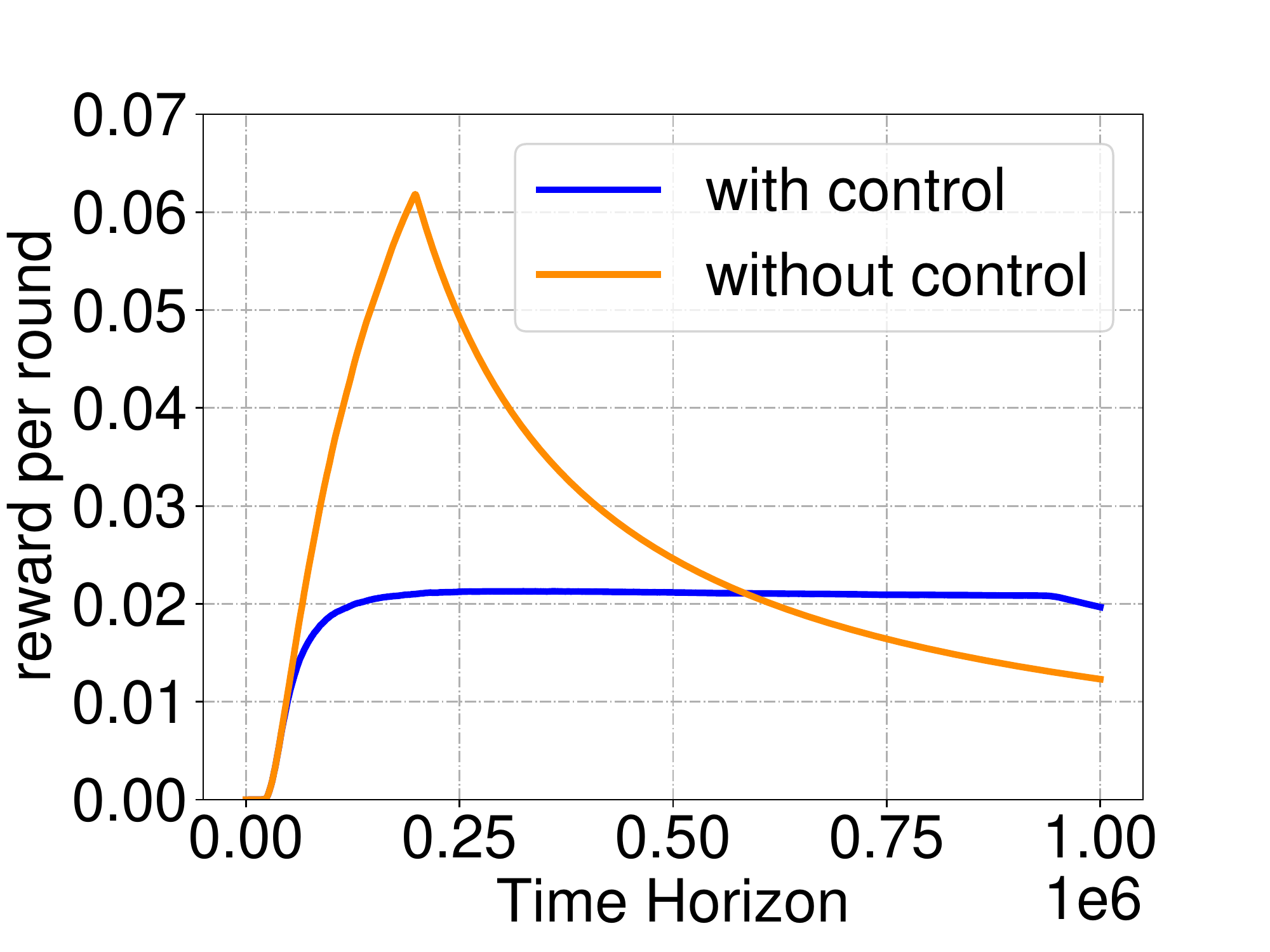}
	\caption{Uniform $v_t$, one-sided feedback.}
	\end{subfigure}
    \caption{Performance of bidding algorithms with and without budget control, under full \textit{(Upper)} and one-sided \textit{(Lower)} feedback, evaluated with respect to the reward per round. In three columns, private values are respectively sampled from: \textit{(Left)} normal distribution $v_t \sim \mathcal{N}(0.6, 0.1)$, \textit{(Middle)} logarithmic normal distribution $\log v_t \sim \mathcal{N}(-0.4, 0.1)$, and \textit{(Right)} uniform distribution $v_t \sim \mathcal{U}(0.25, 1)$.}
    \label{fig: 1}
\end{figure*}

In this section, we empirically evaluate the reward obtained by our proposed algorithms with both full and partial information feedback, using data generated from various distributions. The primary objective of the numerical experiments is to demonstrate the effectiveness of budget management in different settings.
The performance may be further improved by tuning the parameters according to the amount of available budgets. The characterization of such optimal context-dependent parameters is left as an open future problem. 

\paragraph{Setup.} We consider repeated first-price auctions with $T = 10^6$ rounds, budget amount $B = 10^4$ and upper bound on values $\bar{v} = 1$. We generate the sequence of competing bids by sampling each $d_t$ \textit{i.i.d.} from normal distribution $\mathcal{N}(0.4, 0.1)$. 
For the sequence of private values, we consider normal distribution, logarithmic normal distribution and uniform distribution respectively.
Detailed parameters of the private value distributions can be found in \Cref{fig: 1}.

In each experiment, we simulate $T$ rounds auctions under both full and one-sided information structure, and compare our proposed algorithm with ones without budget management, i.e., all the same except for omitting multiplier $\lambda_t$ and using true value $v_t$ instead of $v_t/(1+\lambda_t)$. The performance is evaluated by observing and plotting $\sum_{s=1}^t r_s/t$, the \emph{reward per round} as a function of $t$. For all algorithms, we uniformly set $M = K = 100$, failure probability $\delta = 0.01$, and adopt fixed step size $\epsilon = 1/\sqrt{T}$. For each of the graph we take the average of $20$ independent repetitions of the process.

In the above setup, readers may think $\rho/\bar{v} = 0.01$ is a too tight constraint. However, with an example we show that a seemingly ``tight'' constraint is necessary for budget management to be of even the least use. 
\begin{example}
In $T$-round repeated first price auctions with $v_t, d_t \sim \mathcal{U}(0,1)$, when adopting the optimal strategy for Problem~\eqref{eqn: soft problem}, the soft budget constraint is not binding if $\rho \geq 1/12$.
\end{example}

We note that the study on budget management is only needed when budget is relatively tight, such as $\rho < 1/12$ in the above example. Otherwise, a bidder can simply ``forget'' $B$, adopt unconstrained strategies, without expecting her budget to run out. 

\paragraph{Results and Discussions.} 
The results of three parallel experiments are plotted in \Cref{fig: 1}, with both full and one-sided information feedback considered. Notably, in all instances, our proposed algorithm outperforms the one with no budget control, with respect to the total reward. The latter algorithm gains remarkable rewards in the beginning rounds, yet tends to deplete its budget in an early phase. The reward per round is then inversely proportional to $t$, and is eventually exceeded by algorithms with budget control. Meanwhile, for algorithms with budget control, the budget can also be depleted in some instances, but only at the very ending phase, with a delicate turning in the tail of each curve. This coincides our argument that the algorithm has its expected time of budget depletion close to $T$. We also note that for the bidding algorithm under one-sided information feedback with budget control, $\sum_{s=1}^t r_s/t$ holds steady in most rounds, which indicates that the algorithm manages to achieve stable per-round gain as the budget is diminishing. This further demonstrates the effectiveness of the proposed algorithms on budget management.

\paragraph{Further Experiments on \Cref{lem: main}.} 
We notice that the proposed algorithm performs well in the third experiment where $v_t \sim \mathcal{U}(0.25, 1)$, which does not satisfies the conditions of \Cref{asm: main}. This indicates that the proposed bidding algorithm with one-sided feedback might perform well on a broader class of private value distributions beyond the requirement of \Cref{asm: main}. The following experiment provides numerical evidence that \Cref{lem: main} is very likely to hold for the value distribution $\mathcal{U}(0.25, 1)$. 

\begin{figure}[h]
    \includegraphics[width=0.9\linewidth]{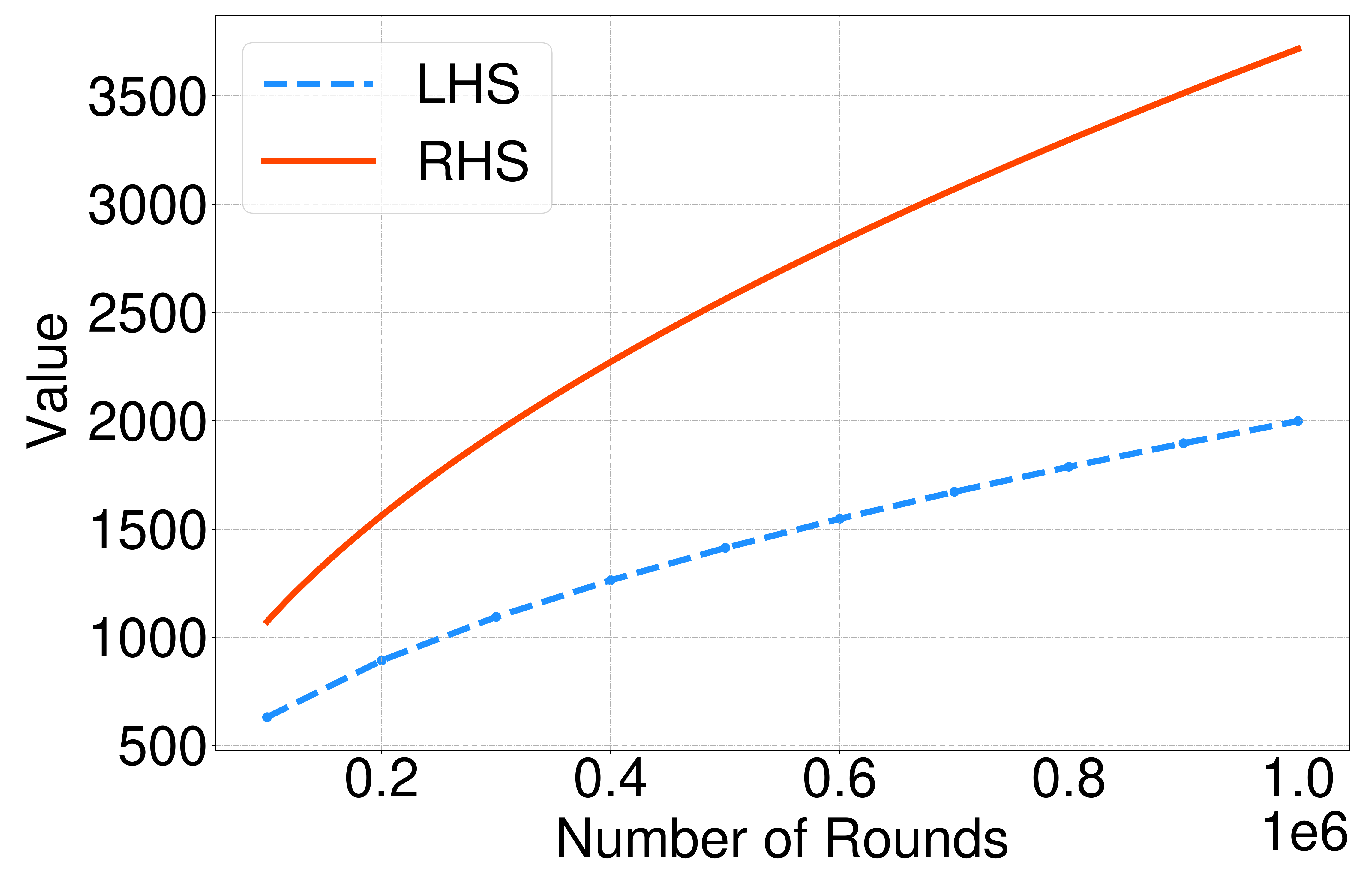}
    \caption{Numerical evidence for \Cref{lem: main} by showing $\mathbb{E} \left[\sum_{t=2}^T ({N_t^{m(t)}})^{-1/2}\right] \leq \sqrt{T\ln T}$, in a case where \Cref{asm: main} does not hold.}
    \label{fig: 2}
\end{figure}

We simulate the bidding algorithm with budget control under one-sided feedback with different time horizons $T = 10^5 \tau$ where $ \tau = 1, \cdots, 10$, while fixing $K = M = 100$. For each horizon $T$, we observe the sequence of values and bids to compute $\sum_{t=2}^T ({N_t^{m(t)}})^{-1/2}$. We repeat the process $10$ times to estimate its expectation, which is compared with $\sqrt{T\ln T}$. The results are plotted in \Cref{fig: 2}.

In \Cref{fig: 2}, the growth rate of the blue curve is notably smaller than $\sqrt{T\ln T}$, supporting the inequality in \Cref{lem: main}. We further conjecture that similar properties may hold for a larger class of distributions, only a subset of which is captured by \Cref{asm: main}. Preciser theoretical characterization of the distribution class is an interesting open problem, which we left as a future direction.
\section{Conclusion}

In this paper, we study design of bidding algorithms for repeated first-prices auctions with budgets, in both full and one-sided feedback models. On the theoretical side, we prove that \cref{algo: full} can achieve an $\widetilde{O}(\sqrt{T})$ regret in the case with full information feedback and that with a technical assumption \Cref{algo: one} can achieve an $\widetilde{O}(\sqrt{T})$ regret in the case with one-sided information feedback. On the practical side, we show that our algorithms can attain effective budget management as well as good performance. The experiments under different distributions provide evidence that our algorithms can be widely applicable.



\newpage
\bibliography{ref}

\begin{thebibliography}{26}
\providecommand{\natexlab}[1]{#1}
\providecommand{\url}[1]{\texttt{#1}}
\expandafter\ifx\csname urlstyle\endcsname\relax
  \providecommand{\doi}[1]{doi: #1}\else
  \providecommand{\doi}{doi: \begingroup \urlstyle{rm}\Url}\fi

\bibitem[Ai et~al.(2022)Ai, Wang, Li, Zhang, Huang, and Deng]{ai2022no}
Ai, R., Wang, C., Li, C., Zhang, J., Huang, W., and Deng, X.
\newblock No-regret learning in repeated first-price auctions with budget
  constraints.
\newblock \emph{arXiv preprint arXiv:2205.14572}, 2022.

\bibitem[Badanidiyuru et~al.(2021)Badanidiyuru, Feng, and
  Guruganesh]{badanidiyuru2021learning}
Badanidiyuru, A., Feng, Z., and Guruganesh, G.
\newblock Learning to bid in contextual first price auctions.
\newblock \emph{arXiv preprint arXiv:2109.03173}, 2021.

\bibitem[Balseiro et~al.(2019)Balseiro, Golrezaei, Mahdian, Mirrokni, and
  Schneider]{balseiro2019contextual}
Balseiro, S., Golrezaei, N., Mahdian, M., Mirrokni, V., and Schneider, J.
\newblock Contextual bandits with cross-learning.
\newblock \emph{Advances in Neural Information Processing Systems}, 32, 2019.

\bibitem[Balseiro \& Gur(2019)Balseiro and Gur]{balseiro2019learning}
Balseiro, S.~R. and Gur, Y.
\newblock Learning in repeated auctions with budgets: Regret minimization and
  equilibrium.
\newblock \emph{Management Science}, 65\penalty0 (9):\penalty0 3952--3968,
  2019.

\bibitem[Balseiro et~al.(2015)Balseiro, Besbes, and
  Weintraub]{balseiro2015repeated}
Balseiro, S.~R., Besbes, O., and Weintraub, G.~Y.
\newblock Repeated auctions with budgets in ad exchanges: Approximations and
  design.
\newblock \emph{Management Science}, 61\penalty0 (4):\penalty0 864--884, 2015.

\bibitem[Balseiro et~al.(2022{\natexlab{a}})Balseiro, Kroer, and
  Kumar]{balseiro2022contextual}
Balseiro, S.~R., Kroer, C., and Kumar, R.
\newblock Contextual standard auctions with budgets: Revenue equivalence and
  efficiency guarantees.
\newblock In \emph{Proceedings of the 23rd ACM Conference on Economics and
  Computation}, pp.\  476--476, 2022{\natexlab{a}}.

\bibitem[Balseiro et~al.(2022{\natexlab{b}})Balseiro, Lu, and
  Mirrokni]{balseiro2022best}
Balseiro, S.~R., Lu, H., and Mirrokni, V.
\newblock The best of many worlds: Dual mirror descent for online allocation
  problems.
\newblock \emph{Operations Research}, 2022{\natexlab{b}}.

\bibitem[Bercu \& Touati(2008)Bercu and Touati]{bercu2008exponential}
Bercu, B. and Touati, A.
\newblock Exponential inequalities for self-normalized martingales with
  applications.
\newblock \emph{The Annals of Applied Probability}, 18\penalty0 (5):\penalty0
  1848--1869, 2008.

\bibitem[Bigler(2019)]{google2019fpa}
Bigler, J.
\newblock Rolling out first price auctions to google ad manager partners.
\newblock
  \url{https://www.blog.google/products/admanager/rolling-out-first-price-auctions-google-ad-manager-partners},
  2019.
\newblock Accessed: 08/01/2023.

\bibitem[Capen et~al.(1971)Capen, Clapp, and Campbell]{capen1971competitive}
Capen, E.~C., Clapp, R.~V., and Campbell, W.~M.
\newblock Competitive bidding in high-risk situations.
\newblock \emph{Journal of petroleum technology}, 23\penalty0 (06):\penalty0
  641--653, 1971.

\bibitem[Chen et~al.(2022)Chen, Wang, Wang, Pan, Shi, Tang, Cai, Ren, Zhu, and
  Deng]{chen2022dynamic}
Chen, Z., Wang, C., Wang, Q., Pan, Y., Shi, Z., Tang, C., Cai, Z., Ren, Y.,
  Zhu, Z., and Deng, X.
\newblock Dynamic budget throttling in repeated second-price auctions.
\newblock \emph{arXiv preprint arXiv:2207.04690}, 2022.

\bibitem[de~la Pena et~al.(2004)de~la Pena, Klass, and Lai]{de2004self}
de~la Pena, V.~H., Klass, M.~J., and Lai, T.~L.
\newblock Self-normalized processes: exponential inequalities, moment bounds
  and iterated logarithm laws.
\newblock \emph{Annals of probability}, pp.\  1902--1933, 2004.

\bibitem[Despotakis et~al.(2021)Despotakis, Ravi, and
  Sayedi]{despotakis2021first}
Despotakis, S., Ravi, R., and Sayedi, A.
\newblock First-price auctions in online display advertising.
\newblock \emph{Journal of Marketing Research}, 58\penalty0 (5):\penalty0
  888--907, 2021.

\bibitem[Esponda(2008)]{esponda2008information}
Esponda, I.
\newblock Information feedback in first price auctions.
\newblock \emph{The RAND Journal of Economics}, 39\penalty0 (2):\penalty0
  491--508, 2008.

\bibitem[Feng et~al.(2022)Feng, Padmanabhan, and Wang]{feng2022online}
Feng, Z., Padmanabhan, S., and Wang, D.
\newblock Online bidding algorithms for return-on-spend constrained
  advertisers.
\newblock \emph{arXiv preprint arXiv:2208.13713}, 2022.

\bibitem[Golrezaei et~al.(2021)Golrezaei, Jaillet, Liang, and
  Mirrokni]{golrezaei2021bidding}
Golrezaei, N., Jaillet, P., Liang, J. C.~N., and Mirrokni, V.
\newblock Bidding and pricing in budget and roi constrained markets.
\newblock \emph{arXiv preprint arXiv:2107.07725}, 2021.

\bibitem[{Google Ad Exchange}(2022)]{google2022adshelp}
{Google Ad Exchange}.
\newblock Bid data sharing.
\newblock \url{https://support.google.com/authorizedbuyers/answer/2696468},
  2022.
\newblock Accessed: 08/01/2023.

\bibitem[Han et~al.(2020{\natexlab{a}})Han, Zhou, Flores, Ordentlich, and
  Weissman]{han2020learning}
Han, Y., Zhou, Z., Flores, A., Ordentlich, E., and Weissman, T.
\newblock Learning to bid optimally and efficiently in adversarial first-price
  auctions.
\newblock \emph{arXiv preprint arXiv:2007.04568}, 2020{\natexlab{a}}.

\bibitem[Han et~al.(2020{\natexlab{b}})Han, Zhou, and Weissman]{han2020optimal}
Han, Y., Zhou, Z., and Weissman, T.
\newblock Optimal no-regret learning in repeated first-price auctions.
\newblock \emph{arXiv preprint arXiv:2003.09795}, 2020{\natexlab{b}}.

\bibitem[Iyer et~al.(2014)Iyer, Johari, and Sundararajan]{iyer2014mean}
Iyer, K., Johari, R., and Sundararajan, M.
\newblock Mean field equilibria of dynamic auctions with learning.
\newblock \emph{Management Science}, 60\penalty0 (12):\penalty0 2949--2970,
  2014.

\bibitem[Klemperer(2004)]{klemperer2004auctions}
Klemperer, P.
\newblock \emph{Auctions: theory and practice}.
\newblock Princeton University Press, 2004.

\bibitem[\lowercase{e}Marketer(2022)]{adspending2022worldwide}
\lowercase{e}Marketer.
\newblock Worldwide ad spending 2022.
\newblock
  \url{https://www.insiderintelligence.com/content/worldwide-ad-spending-2022},
  2022.
\newblock Accessed: 14/12/2022.

\bibitem[Lucking-Reiley(2000)]{lucking2000vickrey}
Lucking-Reiley, D.
\newblock Vickrey auctions in practice: From nineteenth-century philately to
  twenty-first-century e-commerce.
\newblock \emph{Journal of economic perspectives}, 14\penalty0 (3):\penalty0
  183--192, 2000.

\bibitem[Lucking-Reiley et~al.(2007)Lucking-Reiley, Bryan, Prasad, and
  Reeves]{lucking2007pennies}
Lucking-Reiley, D., Bryan, D., Prasad, N., and Reeves, D.
\newblock Pennies from ebay: The determinants of price in online auctions.
\newblock \emph{The journal of industrial economics}, 55\penalty0 (2):\penalty0
  223--233, 2007.

\bibitem[Massart(1990)]{massart1990tight}
Massart, P.
\newblock The tight constant in the dvoretzky-kiefer-wolfowitz inequality.
\newblock \emph{The annals of Probability}, pp.\  1269--1283, 1990.

\bibitem[Zhang et~al.(2022)Zhang, Han, Zhou, Flores, and
  Weissman]{zhang2022leveraging}
Zhang, W., Han, Y., Zhou, Z., Flores, A., and Weissman, T.
\newblock Leveraging the hints: Adaptive bidding in repeated first-price
  auctions.
\newblock \emph{arXiv preprint arXiv:2211.06358}, 2022.

\end{thebibliography}
\bibliographystyle{icml2023}

\newpage
\appendix
\onecolumn
\section{Missing Proofs in \Cref{sec: full}}

\subsection{Proof of \Cref{lem: full dkw}}
Note that \Cref{algo: full estimate} in \Cref{algo: full} essentially estimates $r(v_t, b)$ and $c(b)$ using an empirical distribution $\widetilde{G}_t$:
\begin{align*}
    \widetilde{G}_t(b) = \frac{1}{t-1}\sum_{s=1}^{t-1}\bm{1}\left\{b \geq d_s\right\}.
\end{align*}
By Dvoretzky–Kiefer–Wolfowitz (DKW) inequality \cite{massart1990tight}, we have
\begin{align*}
    {\rm{Pr}}\left(\sup_b |\widetilde{G}_t(b) - G(b)| \geq \sqrt{\frac{\ln \left(2T/\delta\right)}{2(t-1)}} \right) \leq \frac{\delta}{T}.
\end{align*}
Thus with probability at least $1 - \delta$, we have for all $t \geq 2$, 
\begin{align*}
    |\widetilde{r}_t(v_t, b) - r(v_t, b)| \leq & |v_t - b|\cdot |G(b)-\widetilde{G}_t(b)| \leq \bar{v}\cdot \sqrt{\frac{\ln \left(2T/\delta\right)}{2(t-1)}},\\
    |\widetilde{c}_t(b) - c(b)| \leq & |b|\cdot |G(b)-\widetilde{G}_t(b)| \leq \bar{v}\cdot \sqrt{\frac{\ln \left(2T/\delta\right)}{2(t-1)}}.
\end{align*}

\subsection{Proof of \Cref{thm: full}}

We denote by $\tau \coloneqq \sup\{t \leq T: B_{t} \geq \bar{v}\}$ the latest period in which the bidder's remaining budget is larger than her maximum private value under \Cref{algo: full}. We consider an alternative framework in which the bidder is allowed to bid even after budget depletion. Note that the performance of $\pi$ in both the original and alternative frameworks coincide up to time $\tau$. Therefore, 
\begin{align}
    R(\pi) = \bbE_{\vv, \vd} \left[\sum_{t=2}^{\tau} r_t\right] \geq \bbE_{\vv, \vd} \left[\sum_{t=2}^{T} r_t\right] - \bar{v}\cdot \bbE_{\vv, \vd} \left[T - \tau \right]. \label{eqn: full 0}
\end{align}
The inequality holds since $r_t \leq v_t \leq \bar{v}$. Here $r_t$ refers to the reward in the alternate framework where the bidder does not break the loop even if $B_{t+1} < \bar{v}$.

We first characterize the optimal strategy for Problem~\eqref{eqn: soft problem}. The proof of \Cref{lem: full strong} is deferred to \Cref{sec: full strong}.
\begin{lemma}
\label{lem: full strong}
There exists an optimal bidding strategy for Problem~\eqref{eqn: soft problem} that maps the value in each round to a random bid, discarding all historical information. Denoting the optimal bidding strategy by $\alpha^*$, we have
\begin{align*}
    R(\alpha^*) = T\cdot \bbE^{\alpha^*}_{v\sim F} \left[r(v, \alpha^*(v))\right], \quad \bbE^{\alpha^*}_{v\sim F} \left[c(\alpha^*(v))\right] \leq \rho.
\end{align*}
\end{lemma}

Next we start to lower bound the performance of our strategy. For the first term in the right hand side of \eqref{eqn: full 0},  we observe that
\begin{align}
    \bbE^{\pi}_{\vv, \vd} \left[\sum_{t=2}^T r_t\right] = \sum_{t=2}^T \bbE^{\pi}_{\calH_t, v_t} \left[ r(v_t, b_t)\right] = \sum_{t=2}^T \bbE^{\pi}_{\vv, \vd} \left[ r(v_t, b_t)\right] = \bbE^{\pi}_{\vv, \vd} \left[\sum_{t=2}^T  r(v_t, b_t)\right]. \label{eqn: full expect}
\end{align}

Let $\alpha^*$ be the optimal bidding strategy characterized in \Cref{lem: full strong}. By the choice of $b_t$, we have
\begin{align*}
\widetilde{r}_t(v_t, b_t) - \lambda_t \widetilde{c}_t(b_t) \geq \widetilde{r}_t(v_t, \alpha^*(v_t)) - \lambda_t \widetilde{c}_t(\alpha^*(v_t)) 
\end{align*}

Then according to \Cref{lem: full dkw}, with probability at least $1-\delta$, for all $t\geq 2$,
\begin{align*}
r(v_t, b_t) - \lambda_t \widetilde{c}_t(b_t) \geq r(v_t, \alpha^*(v_t)) - \lambda_t c(\alpha^*(v_t)) 
     - (2+\lambda_t) \bar{v}\cdot \sqrt{\frac{\ln \left(2T/\delta\right)}{2(t-1)}}.
\end{align*}
Reordering terms and summing up from $t = 2$ to $T$, we have
\begin{align*}
    \sum_{t=2}^T r(v_t, b_t) \geq \sum_{t=2}^T r(v_t, \alpha^*(v_t)) - \sum_{t=2}^T \lambda_t c(\alpha^*(v_t))  + \sum_{t=2}^T \lambda_t \widetilde{c}_t(b_t) - \sum_{t=2}^T(2+\lambda_t) \bar{v}\cdot \sqrt{\frac{\ln \left(2T/\delta\right)}{2(t-1)}}.
\end{align*}
Notice that the right hand side is upper bounded by $2(T-1)\bar{v}$ since $r(v_t, \alpha^*(v_t)) \leq \bar{v}$ and $\widetilde{c}_t(b_t) \leq 1/(1+\lambda_t)$ by \eqref{eqn: full ogd 2}.

Taking expectations, we obtain
\begin{align*}
    \bbE_{\vv, \vd} \left[\sum_{t=2}^T  r(v_t, b_t)\right] \overset{(\text a)}{\geq} & \bbE^{\alpha^*}_{\vv, \vd} \left[\sum_{t=2}^T r(v_t, \alpha^*(v_t))\right] - \bbE^{\alpha^*}_{\vv, \vd} \left[\sum_{t=2}^T \lambda_t c(\alpha^*(v_t))\right]\\
     & + \bbE_{\vv, \vd} \left[\sum_{t=2}^T \lambda_t \widetilde{c}_t(b_t)\right] - \left(\frac{\bar{v}^2}{\rho} + \bar{v}\right) \sqrt{2T\ln \left(2T/\delta\right)} - \delta \cdot 2(T-1)\bar{v}. \\
    \overset{(\text b)}{\geq} & R(\alpha^*) - \bar{v} - \bbE_{\vv, \vd} \left[\sum_{t=2}^T \lambda_t \left(\rho - \widetilde{c}_t(b_t)\right)\right] - \left(\frac{\bar{v}^2}{\rho} + \bar{v}\right) \sqrt{2T\ln \left(2T/\delta\right)} - \delta \cdot 2(T-1)\bar{v},
\end{align*}
where (a) follows from $\lambda_t \leq \bar{v}/\rho - 1$ by \Cref{lem: full ogd} and (b) follows from \Cref{lem: full strong}.

We now apply a standard analysis of the online gradient descent method to show that the sequence of $\lambda_t$ is not much worse than a hindsight $\lambda$ with respect to gain function $h_t(\lambda_t) = \lambda_t\left(\widetilde{c}_t(b) - \rho\right)$. The proof of \Cref{lem: full ogd} is deferred to \Cref{sec: full ogd}.

\begin{lemma}
\label{lem: full ogd}
For all $t \geq 2$, we have $\lambda_t \in [0, \bar{v}/\rho - 1]$. Moreover, for any $\lambda > 0$, we have
\begin{align}
\label{eqn: full ogd}
\sum_{t=2}^{T} \left(\lambda_{t} - \lambda\right) \left(\rho - \widetilde{c}_t(b_t)\right) \leq \frac{\lambda^2}{2\epsilon} + \frac{(T-1) \epsilon \bar{v}^2}{2}.
\end{align}
\end{lemma}

By using \Cref{lem: full ogd} with $\lambda = 0$, we obtain
\begin{align}
    \bbE_{\vv, \vd} \left[\sum_{t=2}^T  r(v_t, b_t)\right] \geq & R(\alpha^*) - \bar{v} - \bbE_{\vv, \vd} \left[\sum_{t=2}^T \lambda_t \left(\rho - \widetilde{c}_t(b_t)\right)\right] - \left(\frac{\bar{v}^2}{\rho} + \bar{v}\right) \sqrt{2T\ln \left(2T/\delta\right)} - \delta \cdot 2(T-1)\bar{v} \nonumber \\
    \geq & R(\alpha^*) - \bar{v} - \frac{(T-1) \epsilon \bar{v}^2}{2} - \left(\frac{\bar{v}^2}{\rho} + \bar{v}\right) \sqrt{2T\ln \left(2T/\delta\right)} - \delta \cdot 2(T-1)\bar{v}, \label{eqn: full 1}
\end{align}

For the second term in the right hand side of \eqref{eqn: full 0}, we show that the stopping time $\tau$ is close to $T$. The proof of \Cref{lem: full stop} is deferred to \Cref{sec: full stop}
\begin{lemma}
\label{lem: full stop}
For \Cref{algo: full}, with probability at least $1 - 2\delta$, we have
\begin{align*}
T - \tau \leq \frac{\bar{v}}{\rho}\cdot \left(\frac{1}{\epsilon\rho} + \sqrt{2T\ln \left(2T/\delta\right)} + \sqrt{2T\ln \left(1/\delta\right)}\right).
\end{align*}
\end{lemma}

By \Cref{lem: full stop}, we have
\begin{align}
    \bbE_{\vv, \vd} \left[T - \tau \right]\leq & (1-2\delta)\cdot \frac{\bar{v}}{\rho}\cdot \left(\frac{1}{\epsilon\rho} + \sqrt{2T\ln \left(2T/\delta\right)} + \sqrt{2T\ln \left(1/\delta\right)}\right) + 2\delta\cdot T. \label{eqn: full 2}
\end{align}
Plugging \eqref{eqn: full 1} and \eqref{eqn: full 2} into \eqref{eqn: full 0}, we obtain
\begin{align*}
    R(\pi) \geq & R(\alpha^*) - \bar{v} - \frac{(T-1) \epsilon \bar{v}^2}{2}  - \left(\frac{\bar{v}^2}{\rho} + \bar{v}\right) \sqrt{2T\ln \left(2T/\delta\right)} - \delta \cdot 2(T-1)\bar{v} \\
      & - (1-2\delta)\cdot \frac{\bar{v}^2}{\rho}\cdot \left(\frac{1}{\epsilon\rho} + \sqrt{2T\ln \left(2T/\delta\right)} + \sqrt{2T\ln \left(1/\delta\right)}\right) - 2\delta\cdot T\bar{v}.
\end{align*}
By setting the step size to $\epsilon \sim T^{-1/2}$ and the failure probability to $\delta \sim T^{-1}$, \Cref{algo: full} can obtain a regret of order $O(\sqrt{T\log T})$.


\subsection{Proof of \Cref{lem: full strong}}
\label{sec: full strong}
Let $\alpha: [0, \bar{v}] \mapsto \Delta [0, \bar{v}]$ be a bidding strategy that maps $v_t$ to a distribution over $[0, \bar{v}]$. Consider the following optimization problem:
\begin{equation*}
    \begin{aligned}
    \max_{\alpha} & \ \bbE^{\alpha}_{\vv, \vd} \left[\sum_{t=1}^T \bm{1}\left\{\alpha(v_t) \geq d_{t}\right\}\left(v_t - \alpha(v_t)\right)\right] \\
    \text{s.t. } & \ \bbE^{\alpha}_{\vv, \vd} \left[\sum_{t=1}^T \bm{1}\left\{\alpha(v_t) \geq d_{t}\right\}\alpha(v_t)\right] \leq \rho T.
    \end{aligned}
\end{equation*}
Because the sequences of $v_t$ and $d_t$ are independent samples, the problem can be simplified as
\begin{equation}
\label{eqn: simple problem}
    \begin{aligned}
    \max_{\alpha} & \ T\cdot \bbE^{\alpha}_{v\sim F} \left[\left(v - \alpha(v)\right)G(\alpha(v))\right] \\
    \text{s.t.} & \ \bbE^{\alpha}_{v\sim F} \left[\alpha(v)G(\alpha(v))\right] \leq \rho.
    \end{aligned}
\end{equation}
Let $\calA_1$ be the set of all feasible solutions to Problem~\eqref{eqn: simple problem}. Note that this is a convex optimization problem where Slater's condition holds (always bidding $0$ is an interior point). As a result, strong duality holds:
\begin{align*}
    \max_{\alpha \in \calA_1} R(\alpha) = & \min_{\lambda\geq 0} T \cdot \left(\bbE_{v} \left[\max_\alpha \left(v - (1+\lambda)\alpha(v)\right)G\left(\alpha(v)\right)\right] + \lambda\rho \right). 
\end{align*}

The optimal bidding strategy $\alpha^*$ maps $v$ to a distribution over the bids that maximize $\left(v - (1+\lambda^*)b\right)G\left(b\right)$, where $\lambda^*$ satisfies the complementary conditions
\begin{align*}
    \lambda^*\geq 0 \perp \bbE^{\alpha^*}_{v\sim F} \left[\alpha^*(v)G(\alpha^*(v))\right] \leq \rho.
\end{align*}
On the one hand, we have $\calA_1 \subseteq \Pi_1$ so $\max_{\alpha \in \calA_1} R(\alpha) \leq \max_{\pi \in \Pi_1} R(\pi)$. On the other hand, the performance of strategy $\alpha^*$ achieves the right hand side of \eqref{eqn: soft problem upper bound}, an upper bound of $\max_{\pi \in \Pi_1} R(\pi)$. Therefore, $\alpha^*$ is also an optimal strategy for Problem~\eqref{eqn: soft problem}.

\subsection{Proof of \Cref{lem: full ogd}}
\label{sec: full ogd}
By the choice of bid in \Cref{algo: full bid}, we have
\begin{align}
\label{eqn: full ogd 1}
    \widetilde{r}_t(v_t, b_t) - \lambda_t \widetilde{c}_t(b_t) \geq \widetilde{r}_t(v_t, 0) - \lambda_t \widetilde{c}_t(0) \geq 0,
\end{align}
and then,
\begin{align}
    (1+\lambda_t)\widetilde{c}_t(b_t) \leq \widetilde{r}_t(v_t, b_t) + \widetilde{c}_t(b_t) \leq & \frac{1}{t-1}\sum_{s = 1}^{t-1}\bm{1}\left\{b \geq d_s\right\}v_t \leq v_t \leq \bar{v} \nonumber \\
    \Longrightarrow \widetilde{c}_t(b_t) \leq & \frac{\bar{v}}{1+\lambda_t}.
    \label{eqn: full ogd 2}
\end{align}
Meanwhile, inequality \eqref{eqn: full ogd 1} implies $b_t \leq v_t/(1+\lambda_t)$, otherwise $\widetilde{r}_t(v_t, b_t) - \lambda_t \widetilde{c}_t(b_t) \leq 0 \leq \widetilde{r}_t(v_t, 0) - \lambda_t \widetilde{c}_t(0)$, which means that the algorithm should have chosen a smaller bid instead.

According to the update rule of $\lambda_t$, when $\lambda_t \leq \bar{v}/\rho - 1$, we have
\begin{align*}
    \lambda_t - \epsilon\left(\rho - \widetilde{c}_t(b_t)\right) \overset{(\text a)}{\leq} \lambda_t + \frac{\epsilon \bar{v}}{1 + \lambda_t} - \epsilon\rho \overset{(\text b)}{\leq} \max\{\epsilon \bar{v} - \epsilon \rho, \frac{\bar{v}}{\rho} - 1\}
    \overset{(\text c)}{=} \frac{\bar{v}}{\rho} - 1,
\end{align*}
where (a) follows from inequality \eqref{eqn: full ogd 2}, (b) follows from that $\phi(x) = x + \epsilon\bar{v}/(1+x)$ is convex over $\bbR_+$, and (c) holds since $\epsilon = 1/\sqrt{T} < 1/\rho$. Because we take $\lambda_2 = 0$ in initialization, by induction, we have $\lambda_t \in [0, \bar{v}/\rho - 1]$ for all $t \geq 2$.

Again by the update rule of $\lambda_t$ in \Cref{algo: full update}, we have for any $\lambda \geq 0$,
\begin{align*}
    \|\lambda_{t+1} - \lambda\|_2^2 \overset{(\text a)}{\leq} & \|\lambda_{t} - \epsilon (\rho - \widetilde{c}_t(b_t)) - \lambda\|_2^2, \\
    = & \|\lambda_{t} - \lambda\|^2 - 2\epsilon \left(\lambda_{t} - \lambda\right) \left(\rho - \widetilde{c}_t(b_t) \right) + \epsilon^2 \|\rho - \widetilde{c}_t(b_t)\|_2^2 \\
    \overset{(\text b)}{\leq} & \|\lambda_{t} - \lambda\|^2 - 2\epsilon \left(\lambda_{t} - \lambda\right) \left(\rho - \widetilde{c}_t(b_t) \right) + \epsilon^2 \bar{v}^2,
\end{align*}
where (a) follows from a standard contraction property of projection operator and (b) holds by $\rho \leq \bar{v}$ and \eqref{eqn: full ogd 2}.

Reordering terms and summing up from $t = 2$ to $T$, we have
\begin{align*}
    \sum_{t=2}^{T} \left(\lambda_{t} - \lambda\right) \left(\rho - \widetilde{c}_t(b_t)\right) \leq & \frac{\|\lambda_{2} - \lambda\|_2^2 - \|\lambda_{T+1} - \lambda\|_{2}^2}{2\epsilon} + \frac{(T-1) \epsilon \bar{v}^2}{2}.
\end{align*}
which leads to \eqref{eqn: full ogd} with $\lambda_2=0$.

\subsection{Proof of \Cref{lem: full stop}}
\label{sec: full stop}
Reordering $\lambda_{t+1} \geq \lambda_{t} - \epsilon (\rho -\widetilde{c}_t(b_t))$ and summing over $t = 2, \ldots, \tau$, we have
\begin{align}
\label{eqn: full stop 1}
    \sum_{t=2}^{\tau} \left(\widetilde{c}_t(b_t) - \rho\right) & \leq \frac{\lambda_{\tau+1}}{\epsilon} \leq \frac{\bar{v}/\rho - 1}{\epsilon}. 
\end{align}

For the left hand side of \eqref{eqn: full stop 1}, we use inequality \eqref{eqn: full dkw c}. With probability at least $1-\delta$,
\begin{align}
    \sum_{t=2}^{\tau} \left(\widetilde{c}_t(b_t) - \rho\right) \geq \sum_{t=2}^{\tau} c(b_t) - \tau\rho - \sum_{t=2}^{\tau} \bar{v}\cdot \sqrt{\frac{\ln \left(2T/\delta\right)}{2(t-1)}} \label{eqn: full stop 2}
\end{align}

Let $X_t \coloneqq \sum_{s=2}^{t} (c_s - c(b_s))$. Because $\bbE_{d_t} [c_t - c(b_t)|\calH_t] = 0$, we know that$\{X_1, X_2, \ldots, X_{\tau}\}$ is a martingale. Applying Azuma-Hoeffding inequality, we have the following inequality holds with failure probability at most $\delta$, 
\begin{align}
\label{eqn: full stop 3}
    \sum_{t=2}^{\tau} \left(c_t - c(b_t)\right) \leq \bar{v}\cdot \sqrt{2T\ln (1/\delta)}.
\end{align}

According to the definition of $\tau$, when $\tau < T$,
\begin{align}
\label{eqn: full stop 4}
    B_{\tau + 1} < \bar{v} \Longrightarrow \sum_{t=2}^{\tau} c_t > \rho T - \bar{v}.
\end{align}

Combining \eqref{eqn: full stop 1}, \eqref{eqn: full stop 2}, \eqref{eqn: full stop 3} and \eqref{eqn: full stop 4}, we obtain with probability at least $1-2\delta$, 
\begin{align}
\rho(T - \tau) \leq & \frac{\bar{v}/\rho - 1}{\epsilon} + \sum_{t=2}^{\tau} \bar{v}\cdot \sqrt{\frac{\ln \left(2T/\delta\right)}{2(t-1)}} + \bar{v}\cdot \sqrt{2T\ln (1/\delta)} + \bar{v} \nonumber \\
\leq & \bar{v}\cdot \left(\frac{1}{\epsilon\rho} + \sqrt{2T\ln \left(2T/\delta\right)} + \sqrt{2T\ln \left(1/\delta\right)}\right), \label{eqn: full step 3-5}
\end{align}
Note that when $\tau = T$, the inequality holds trivially.
\section{Missing Proofs in \Cref{sec: one}}

\subsection{Proof of \Cref{lem: one azuma}}
By definition, we have
\begin{align*}
   & \widetilde{r}_t(v^m, b^k) - r(v^m, b^k) \\
   = & \left(v^m - b^k\right) \cdot \frac{\sum_{s=1}^{t-1} \bm{1}\left\{b_s \leq b^k\right\}\left(\bm{1}\left\{b^k \geq d_s\right\} - G(b^k)\right)}{\sum_{s=1}^{t-1} \bm{1}\left\{b_s \leq b^k\right\}}.
\end{align*}
We denote the above numerator by $X_t$. As $\bbE [X_{t+1} - X_{t} | \calH_{t}] = 0$, the sequence of $X_t$ is a martingale adapted to the filtration $\left\{\calH_1, \calH_2, \ldots\right\}$. Next we make use the following two lemmas.

\begin{lemma}[\citet{bercu2008exponential}]
\label{lem: self norm 1}
Let $\{X_1, X_2, \ldots\}$ be a locally square integrable martingale. Denote
\begin{align*}
    V_t(\mu) = \exp \left(\mu X_t - \frac{\mu^2}{2}\left(\left<X\right>_t + \left[X\right]_t\right)\right),
\end{align*}
where the predictable quadratic variation $\left<X\right>_t$ and the total quadratic variation $\left[X\right]_t$ are respectively defined by
\begin{align*}
    \left<X\right>_t =  \sum_{s=1}^{t-1} \bbE \left[\left(X_s - X_{s-1}\right)^2|\calH_{s-1}\right], \quad \left[X\right]_t =  \sum_{s=1}^{t-1} \left(X_s - X_{s-1}\right)^2.
\end{align*}
Then, for any $\mu$, the sequence of $V_t(\mu)$ is a positive super-martingale with $\bbE\left[V_t(\mu)\right] \leq 1$.
\end{lemma}

\begin{lemma}[\citet{de2004self}]
\label{lem: self norm 2}
Let $A$ and $B \geq 0$ be two random variables satisfying for any $\mu$, 
\begin{align*}
    \bbE \left[\mu A - \frac{\mu^2}{2}B^2\right] \leq 1.
\end{align*}
The for any $x \geq \sqrt{2}$, $y > 0$, we have
\begin{align*}
    {\rm{Pr}}\left(|A|\bigg/\sqrt{\left(B^2 + y\right) \left(1 + \frac{1}{2}\ln{\left(\frac{B^2}{y} + 1\right)}\right)} \geq x \right) \leq \exp{\left(-\frac{x^2}{2}\right)}
\end{align*}
\end{lemma}

By \Cref{lem: self norm 1}, $A = X_t$ and $B = \sqrt{\left<X\right>_t + \left[X\right]_t}$ satisfies the conditions of \Cref{lem: self norm 2}. Taking $x = \sqrt{2\ln{\left(KT/\delta\right)}}$ and $y=1$, we obtain
\begin{align*}
    {\rm{Pr}}\left(\frac{|A|}{\sqrt{\left(B^2 + 1\right) \left(2 + \ln{\left(B^2 + 1\right)}\right)}} 
    \geq \sqrt{\ln{\left(KT/\delta\right)}} \right) \leq \frac{\delta}{KT}.
\end{align*}
Next, it holds that
\begin{align*}
    \left(B^2 + 1\right) \left(2 + \ln{\left(B^2 + 1\right)}\right) \overset{(\text a)}{\leq} & \left(2n_t^k + 1\right) \left(2 + \ln{\left(2 n_t^k + 1\right)}\right) \\
    \overset{(\text b)}{\leq} & n_t^k \left(6 + 3\ln{\left(2 t - 1\right)}\right) \\
    \overset{(\text c)}{\leq} & 4 n_t^k \ln{T}.
\end{align*}
where (a) holds because $B^2 = \left<X\right>_t + \left[X\right]_t \leq 2 n_t^k$, (b) holds since $n_t^k \in [1, t-1]$ (note that $b_1 = 0$), and (c) holds when $T$ is sufficiently large.

Thus, we have with probability at least $1-\delta$, $\forall t\geq 2, m\in [M], k\in [K]$,
\begin{align*}
   |\widetilde{r}_t(v^m, b^k) - r(v^m, b^k)| \leq \bar{v}\cdot \sqrt{\frac{4 \ln{T}\ln{\left(KT/\delta\right)}}{n_t^k}}.
\end{align*}
The same analysis goes for $\widetilde{c}_t(b^k)$.

\subsection{Proof of \Cref{lem: main}}
We first prove that for any $t \geq 2$, 
\begin{align}
\label{eqn: one count lower bound}
    N_t^{m(t)} \geq 1 + \sum_{s=2}^{t-1} \bm{1}\left\{\frac{v_s}{1+\lambda_s} \leq \frac{v_t}{1+\lambda_t}\right\}.
\end{align}
Note that the bidder always bid $0$ in the first round so $N_t^{m(t)} \geq 1$. For every past round $2 \leq s < t$ with $v_s/(1+\lambda_s) \leq v_t/(1+\lambda_t)$, as $v^{m(s)} \leq v^{m(t)}$, \Cref{algo: one partial order} in \Cref{algo: one} guarantees that $\inf \calB_s^{m(s)} \leq \inf \calB_s^{m(t)}$. Also notice that $\calB_{t-1}^{m(t)} \subseteq \calB_{s}^{m(t)}$ by the bid elimination rule. Therefore, we have for all $b^k \in \calB_{t-1}^{m(t)}$, 
\begin{align}
\label{eqn: one partial order bid}
    b_s = \inf \calB_s^{m(s)} \leq \inf \calB_{t-1}^{m(t)} \leq b^k.
\end{align}
By definition,
\begin{align*}
    N_t^{m(t)} = \min_{b^k\in \calB_{t-1}^{m(t)}} n_t^k = \min_{b^k\in \calB_{t-1}^{m(t)}} \sum_{s=1}^{t-1} \bm{1}\left\{b_s \leq b^k\right\}.
\end{align*}
The inequality \eqref{eqn: one partial order bid} implies that every past value $v_s$ with $v_s/(1+\lambda_s) \leq v_t/(1+\lambda_t)$ has contributed to the value of $N_t^{m(t)}$ by $1$, which leads to the result of \eqref{eqn: one count lower bound}.

Next, due to $\lambda_t \leq \bar{v}/\rho - 1$ by \Cref{lem: full ogd}, we further have $N_t^{m(t)} \geq 1 + \sum_{s=2}^{t-1} \bm{1}\left\{v_s \leq (\rho/\bar{v})v_t\right\}$. Then,
\begin{align*}
    \bbE_{\vv, \vd} \left[\sum_{t=2}^T \sqrt{1\big/N_t^{m(t)}}\right] \leq & \sum_{t=2}^T \bbE_{\vv, \vd} \left[ \sqrt{\frac{1}{1 + \sum_{s=2}^{t-1} \bm{1}\left\{v_s \leq (\rho/\bar{v})v_t\right\}}}\right] \leq \sum_{t=2}^T \sqrt{\bbE_{\vv, \vd} \left[ \frac{1}{1 + \sum_{s=2}^{t-1} \bm{1}\left\{v_s \leq (\rho/\bar{v})v_t\right\}}\right]},
\end{align*}
where the last equality follows from $\bbE[y^2] \geq (\bbE[y])^2$ for any random variable $y$.

Conditioned on $v_t$, the sum $\sum_{s=2}^{t-1} \bm{1}\left\{v_s \leq (\rho/\bar{v})v_t\right\}$ follows a binomial distribution ${\rm{Binomial}}(t-2, F((\rho/\bar{v})v_t))$. Thus,
\begin{align*}
    \bbE_{v_1, \ldots, v_{t-1}} \left[ \frac{1}{1 + \sum_{s=2}^{t-1} \bm{1}\left\{v_s \leq (\rho/\bar{v})v_t\right\}} \right] = & \sum_{s = 0}^{t-2} \frac{1}{1+s} \binom{t-2}{s} F^s((\rho/\bar{v})v_t)\left(1-F((\rho/\bar{v})v_t)\right)^{t-2-s} \\
    = & \sum_{s = 0}^{t-2} \frac{1}{t-1} \binom{t-1}{s+1} F^{s}((\rho/\bar{v})v_t)\left(1-F((\rho/\bar{v})v_t)\right)^{t-2-s} \\
    = & \frac{1}{t-1} \sum_{s = 1}^{t-1} \binom{t-1}{s} F^{s-1}((\rho/\bar{v})v_t)\left(1-F((\rho/\bar{v})v_t)\right)^{t-1-s} \\
    = & \frac{1}{(t-1)F((\rho/\bar{v})v_t)} \left(1 - \left(1-F((\rho/\bar{v})v_t)\right)^{t-1}\right) \\
    \leq & \frac{1}{(t-1)F((\rho/\bar{v})v_t)}.
\end{align*}
Note that the conditional expectation is also upper bounded by $1$ since $\sum_{s=2}^{t-1} \bm{1}\left\{v_s \leq (\rho/\bar{v})v_t\right\} \geq 0$. Consequently, we have
\begin{align*}
    \bbE_{\vv, \vd} \left[ \frac{1}{1 + \sum_{s=2}^{t-1} \bm{1}\left\{v_s \leq (\rho/\bar{v})v_t\right\}} \right] \leq & \bbE_{v_t} \left[ \min\left\{\frac{1}{(t-1)F((\rho/\bar{v})v_t)}, 1\right\} \right] \\
    = & \int_{(t-1)F((\rho/\bar{v})v_t) \geq 1} \frac{1}{(t-1)F((\rho/\bar{v})v_t)} \, dF(v_t) + \int_{(t-1)F((\rho/\bar{v})v_t) \leq 1} \, dF(v_t) \\
    = & \frac{F(v_t)}{(t-1)F((\rho/\bar{v})v_t)} \Bigg|_{(\bar{v}/\rho)F^{-1}(1/(t-1))}^{\bar{v}} \\
    & - \int_{ (\bar{v}/\rho)F^{-1}(1/(t-1))}^{\bar{v}} F(v_t) \, d\left(\frac{1}{(t-1)F((\rho/\bar{v})v_t)}\right) + F((\bar{v}/\rho)F^{-1}(1/(t-1))) \\
    = & \frac{1}{(t-1)F(\rho)} + \int_{(\bar{v}/\rho)F^{-1}(1/(t-1))}^{\bar{v}} \frac{(\rho/\bar{v})F(v_t)f((\rho/\bar{v})v_t)}{(t-1)F^2((\rho/\bar{v})v_t)} \, d\left(v_t\right) \\
    \leq & \frac{1}{(t-1)F(\rho)} + \frac{\overline{f}\bar{v}}{(t-1)\underline{f}\rho}\cdot \ln{F((\rho/\bar{v})v_t)} \bigg|_{(\bar{v}/\rho)F^{-1}(1/(t-1))}^{\bar{v}} \\
    = & \underbrace{\left(\frac{1}{F(\rho)} + \frac{\overline{f}\bar{v}\ln{F(\rho)}}{\underline{f}\rho}\right)}_{C_1}\cdot \frac{1}{t-1} + \underbrace{\frac{\overline{f}\bar{v}}{\underline{f}\rho}}_{C_2}\cdot \frac{\ln{(t-1)}}{t-1} \\
\end{align*}
Summing up the square-roots over $t = 2, \ldots, T$, we have
\begin{align*}
    \bbE_{\vv, \vd} \left[\sum_{t=2}^T \sqrt{1\big/N_t^{m(t)}}\right] \leq & \sum_{t=2}^T \sqrt{C_1 \cdot \frac{1}{t-1} + C_2 \frac{\ln{(t-1)}}{t-1}} \\
    \leq & \sqrt{C_1} \sum_{t=1}^{T-1} \sqrt{\frac{1}{t}} + \sqrt{C_2} \sum_{t=1}^{T-1} \sqrt{\frac{\ln{t}}{t}} \\
    \leq & \left(\sqrt{C_1}+\sqrt{C_2 \ln{T}}\right) \sum_{t=1}^{T-1} t^{-1/2} \\
    \leq & 2\left(\sqrt{C_1}+\sqrt{C_2 \ln{T}}\right)\sqrt{T},
\end{align*}
where the second inequality follows from $\sqrt{x+y} \leq \sqrt{x} + \sqrt{y}$. 

\subsection{Proof of \Cref{thm: one}}
Similar to the proof of \Cref{thm: full}, we denote by $\tau \coloneqq \sup\{t \leq T: B_{t} \geq \bar{v}\}$ the last round before budget depletion under \Cref{algo: one} and consider an alternative framework in which the bidder is allowed to bid even after budget depletion. Then, we have
\begin{align}
    R(\pi) = \bbE_{\vv, \vd} \left[\sum_{t=2}^{\tau} r_t\right] \geq \bbE_{\vv, \vd} \left[\sum_{t=2}^{T} r_t\right] - \bar{v}\cdot \bbE_{\vv, \vd} \left[T - \tau \right] = & \bbE^{\pi}_{\vv, \vd} \left[\sum_{t=2}^T  r(v_t, b_t)\right] - \bar{v}\cdot \bbE_{\vv, \vd} \left[T - \tau \right], \label{eqn: one 0}
\end{align}
where the last equality follows equation \eqref{eqn: full expect}.

First, we make use of the following lemma, the proof of which is deferred to \Cref{sec: one elimination}
\begin{lemma}
\label{lem: one elimination}
Let $\widetilde{b}(v) = \arg\max_{b\in \calB} (v - b)G(b)$ (taking the smallest  $b$ if there are ties). Then for \cref{algo: one}, with probability at least $1-\delta$, $\forall t \geq 2, m\in [M]$, $\widetilde{b}(v^m) \in B_t^m$.
\end{lemma}

Then with probability at least $1 - \delta$, for all $t$, 
\begin{align}
    r(v^{m(t)}, b_t) \geq & \widetilde{r}_t(v^{m(t)}, b_t) - w_t^{m(t)} \nonumber \\
   \geq & \widetilde{r}_t(v^{m(t)}, \widetilde{b}(v^{m(t)})) - 3w_t^{m(t)} \nonumber \\
   \geq & r(v^{m(t)}, \widetilde{b}(v^{m(t)})) - 4w_t^{m(t)}, \label{eqn: one 1}
\end{align}
where the first and third inequalities follows from \Cref{lem: one azuma}, and the second inequality holds by \Cref{lem: one elimination}. 

Next, for the left hand side of \eqref{eqn: one 1}, we have
\begin{align}
    r(v^{m(t)}, b_t) = & (v^{m(t)} - b_t)G(b_t) \nonumber \\
    = & \left(\frac{v_t}{1+\lambda_t} - b_t\right)G(b_t) - \left(\frac{v_t}{1+\lambda_t} - v^{m(t)}\right)G(b_t) \nonumber \\
    = & \frac{1}{1+\lambda_t} \left(r(v_t, b_t) - \lambda_tc(b_t)\right) - \left(\frac{v_t}{1+\lambda_t} - v^{m(t)}\right)G(b_t).
    \label{eqn: one 2}
\end{align}

Let $b^*(v) = \arg\max_b (v-b)G(b)$ (taking the smallest $b$ if there are ties) and $b'(v) = \min \{b \in \calB: b \geq b^*(v)\}$. For the right hand side of \eqref{eqn: one 1}, we have
\begin{align}
    r(v^{m(t)}, \widetilde{b}(v^{m(t)})) \overset{(\text a)}{\geq} & r(v^{m(t)}, b'(v^{m(t)})) \nonumber \\
    = & (v^{m(t)} - b'(v^{m(t)}))G(b'(v^{m(t)})) \nonumber \\
     \overset{(\text b)}{\geq} & (v^{m(t)} - b'(v^{m(t)}))G(b^*(v^{m(t)})) \nonumber\\
     \overset{(\text c)}{\geq} & (v^{m(t)} - b^*(v^{m(t)}))G(b^*(v^{m(t)})) - \frac{\bar{v}}{K} \nonumber \\
    = & r(v^{m(t)}, b^*(v^{m(t)})) - \frac{\bar{v}}{K}, \label{eqn: one 3}
\end{align}
where (a) and (b) hold by the definition of $\widetilde{b}(v)$, and (c) holds since $b'(v) \leq b^*(v) + \bar{v}/K$.

Let $\alpha^*$ be the optimal bidding strategy characterized in \Cref{lem: full strong}. By the definition of $b^*(v)$, We have
\begin{align}
    r(v^{m(t)}, b^*(v^{m(t)})) \geq & r(v^{m(t)}, \alpha^*(v_t)) \nonumber \\
    = & (v^{m(t)} - \alpha^*(v_t))G(\alpha^*(v_t)) \nonumber \\
    = & \left(\frac{v_t}{1+\lambda_t} - \alpha^*(v_t)\right)G(\alpha^*(v_t)) - \left(\frac{v_t}{1+\lambda_t} - v^{m(t)}\right)G(\alpha^*(v_t)) \nonumber \\
    = & \frac{1}{1+\lambda_t} \left(r(v_t, \alpha^*(v_t)) - \lambda_tc(\alpha^*(v_t))\right) - \left(\frac{v_t}{1+\lambda_t} - v^{m(t)}\right)G(\alpha^*(v_t)). \label{eqn: one 4}
\end{align}

Putting \eqref{eqn: one 1}, \eqref{eqn: one 2}, \eqref{eqn: one 3} and \eqref{eqn: one 4} together, we obtain
\begin{align*}
r(v_t, b_t) - \lambda_tc(b_t) \geq & r(v_t, \alpha^*(v_t)) - \lambda_tc(\alpha^*(v_t)) - (1+\lambda_t)\left(\frac{\bar{v}}{K} + 4w_t^m\right) \\
& + (1+\lambda_t)\left(\frac{v_t}{1+\lambda_t} - v^{m(t)}\right)\left(G(b_t) - G(\alpha^*(v_t))\right) \\
\geq & r(v_t, \alpha^*(v_t)) - \lambda_tc(\alpha^*(v_t)) - (1+\lambda_t)\left(\frac{\bar{v}}{K} + \frac{\bar{v}}{M} + 4w_t^{m(t)}\right),
\end{align*}
where the second inequality holds since $v_t/(1+\lambda_t) - v^{m(t)} \leq \bar{v}/M$. Reordering terms and summing up from $t = 2$ to $T$, we have with probability at least $1-\delta$,
\begin{align*}
    \sum_{t=2}^T r(v_t, b_t) \geq & \sum_{t=2}^T r(v_t, \alpha^*(v_t)) - \sum_{t=2}^T \lambda_t c(\alpha^*(v_t))  - \sum_{t=2}^T  (1+\lambda_t)\left(\frac{\bar{v}}{K} + \frac{\bar{v}}{M} + 4w_t^{m(t)}\right) + \sum_{t=2}^T \lambda_t c(b_t) \\
    \geq & \sum_{t=2}^T r(v_t, \alpha^*(v_t)) - \sum_{t=2}^T \lambda_t c(\alpha^*(v_t))  - \sum_{t=2}^T  (1+\lambda_t)\left(\frac{\bar{v}}{K} + \frac{\bar{v}}{M} + 4w_t^{m(t)}\right) + \sum_{t=2}^T \lambda_t \widetilde{c}_t(b_t) - \sum_{t=2}^T \lambda_t w_t^{m(t)}, 
\end{align*}
where we further apply \Cref{lem: one azuma} for the second inequality. 

Taking expectations, we obtain
\begin{align}
    \bbE_{\vv, \vd} \left[\sum_{t=2}^T  r(v_t, b_t)\right] \geq & \bbE^{\alpha^*}_{\vv, \vd} \left[\sum_{t=2}^T r(v_t, \alpha^*(v_t))\right] - \bbE^{\alpha^*}_{\vv, \vd} \left[\sum_{t=2}^T \lambda_t c(\alpha^*(v_t))\right] - \frac{\bar{v}^2}{\rho} (T-1)\left(\frac{1}{K} + \frac{1}{M}\right) \nonumber \\
     & + \bbE_{\vv, \vd} \left[\sum_{t=2}^T \lambda_t \widetilde{c}_t(b_t)\right] - \left(\frac{5\bar{v}}{\rho} - 1\right) \bbE_{\vv, \vd} \left[\sum_{t=2}^T w_t^{m(t)}\right] - \delta \cdot 2(T-1)\bar{v} \nonumber \\
    \overset{(\text a)}{\geq} & R(\alpha^*) - \bar{v} - \bbE_{\vv, \vd} \left[\sum_{t=2}^T \lambda_t \left(\rho - \widetilde{c}_t(b_t)\right)\right] - \frac{\bar{v}^2}{\rho} (T-1)\left(\frac{1}{K} + \frac{1}{M}\right) \nonumber \\
     & - \left(\frac{5\bar{v}}{\rho} - 1\right) \bbE_{\vv, \vd} \left[\sum_{t=2}^T w_t^{m(t)}\right] - \delta \cdot 2(T-1)\bar{v} \nonumber \\
    \overset{(\text a)}{\geq} & R(\alpha^*) - \bar{v} - \frac{(T-1) \epsilon \bar{v}^2}{2} - \frac{\bar{v}^2}{\rho} (T-1)\left(\frac{1}{K} + \frac{1}{M}\right) \nonumber \\
     & - \left(\frac{5\bar{v}}{\rho} - 1\right) \bbE_{\vv, \vd} \left[\sum_{t=2}^T w_t^{m(t)}\right] - \delta \cdot 2(T-1)\bar{v},
     \label{eqn: one 5}
\end{align}
where (a) follows from \Cref{lem: full strong} and $\lambda_t \leq \bar{v}/\rho - 1$ in \Cref{lem: full ogd}, (b) holds by using \Cref{lem: full ogd} with $\lambda = 0$. Remark that \Cref{lem: full ogd} still holds for \Cref{algo: one} under one-sided information feedback.

Next, we provide a result for one-sided information feedback analogous to \Cref{lem: full stop}.
\begin{lemma}
\label{lem: one stop}
For \Cref{algo: one}, with probability at least $1 - 2\delta$, we have
\begin{align*}
T - \tau \leq \frac{\bar{v}}{\rho}\cdot \left(\frac{1}{\epsilon\rho} + \sqrt{2T\ln \left(1/\delta\right)}\right) + \frac{1}{\rho}\bbE_{\vv, \vd} \left[\sum_{t=2}^T w_t^{m(t)}\right].
\end{align*}
\end{lemma}

The proof is exactly the same to that of \Cref{lem: full stop}, except that we use \Cref{lem: one azuma} rather than \Cref{lem: full dkw} to bound the estimation error. We omit the proof.

By \Cref{lem: one stop}, we have
\begin{align}
\bbE_{\vv, \vd} \left[T - \tau\right] \leq & (1-2\delta) \cdot \frac{\bar{v}}{\rho}\cdot \left(\frac{1}{\epsilon\rho} + \sqrt{2T\ln \left(1/\delta\right)}\right) + (1-2\delta) \cdot \frac{1}{\rho}\bbE_{\vv, \vd} \left[\sum_{t=2}^T w_t^{m(t)}\right] + 2\delta \cdot T \label{eqn: one 6}
\end{align}

Plugging \eqref{eqn: one 5} and \eqref{eqn: one 6} into \eqref{eqn: one 0}, we obtain
\begin{align*}
    R(\pi) \geq & R(\alpha^*) - \bar{v} - \frac{(T-1) \epsilon \bar{v}^2}{2} - \frac{\bar{v}^2}{\rho} (T-1)\left(\frac{1}{K} + \frac{1}{M}\right) \\
    & - \delta \cdot 2(T-1)\bar{v}
     - (1-2\delta)\cdot \frac{\bar{v}^2}{\rho}\cdot \left(\frac{1}{\epsilon\rho} + \sqrt{2T\ln \left(1/\delta\right)}\right) - 2\delta\cdot T\bar{v} \\
      & - \left(\frac{5\bar{v}^2}{\rho} - \bar{v} + (1-2\delta)\frac{\bar{v}^2}{\rho}\right)\sqrt{4\ln{T}\log(KT/\delta)} \bbE_{\vv, \vd} \left[\sum_{t=2}^T\sqrt{\frac{1}{N_t^{m(t)}}}\right].
\end{align*}
By setting the step size to $\epsilon \sim T^{-1/2}$ and the failure probability to $\delta \sim T^{-1}$, together with \Cref{lem: main}, we can obtain the desired regret bound.


\subsection{Proof of \Cref{lem: one elimination}}
\label{sec: one elimination}
We first prove that $\widetilde{b}(v)$ is non-decreasing in $v$. Let $v^s$ and $v^m$ be two values with $v^s \leq v^m$. Then for any $b\in \calB$ with $b \geq \widetilde{b}(v^m)$, we have
\begin{align*}
    r(v^s, b) = & (v^s - b)G(b) = (v^m - b)G(b) - (v^m - v^s)G(b) \\
    \leq & (v^m - \widetilde{b}(v^m))G(b^*(v^m)) - (v^m - v^s)G(\widetilde{b}(v^m)) \\
    = & (v^s - \widetilde{b}(v^m)) G(\widetilde{b}(v^m)) = r(v^s, \widetilde{b}(v^m)).
\end{align*}
Thus, it holds that $\widetilde{b}(v^s) \leq \widetilde{b}(v^m)$ by the definition of $\widetilde{b}(v)$. 

Next, we prove the result by induction. Suppose that for all $s < m$, $\widetilde{b}(v^s) \in B_{t}^s$ in round $t$. By the non-decreasing property of $\widetilde{b}(v)$, we have
\begin{align*}
    \widetilde{b}(v^m) \geq \widetilde{b}(v^s) \geq \inf B_{t}^s.
\end{align*}
Thus, $\widetilde{b}(v^m)$ is not eliminated by \Cref{algo: one partial order}.
As for \Cref{algo: one high reward bid}, we have for any $k \in \calB_{t-1}^m$, 
\begin{align*}
    \widetilde{r}_t(v^m, \widetilde{b}(v^m)) \overset{(\text a)}{\geq} & r(v^m, \widetilde{b}(v^m)) - w_{t}^m \overset{(\text b)}{\geq} r(v^m, b^k) - w_{t}^m \overset{(\text c)}{\geq} \widetilde{r}_t(v^m, b^k) - 2w_{t}^m,
\end{align*}
where (a) and (c) follows from \Cref{lem: one azuma}, and (b) holds by the definition of $\widetilde{b}(v)$. Thus, $\widetilde{b}(v^m)$ is also not eliminated by \Cref{algo: one high reward bid} for all $t$ and $m$ with probability at least $1-\delta$, which finishes the proof.


\end{document}